\documentclass[fleqn,usenatbib]{mnras}

\usepackage{newtxtext,newtxmath,pdflscape}

\usepackage[T1]{fontenc}
\usepackage{ae,aecompl}

\usepackage{graphicx}	
\usepackage{amsmath}	
\usepackage{ulem}
\usepackage{soul}

\def\apj{ApJ}
\def\apjl{ApJL}
\def\aap{A\&A}
\def\mnras{MNRAS}
\def\pasp{PASP}
\def\araa{ARAA}
\def\aj{AJ}
\def\aapr{A\&A Rev.}
\def\apjs{ApJ Supp}

\def\nat{Natur}

\newcommand{\lsim }{{\lower0.8ex\hbox{$\buildrel <\over\sim$}}}
\newcommand{\gsim }{{\lower0.8ex\hbox{$\buildrel >\over\sim$}}}
\newcommand{\Msun}{\ifmmode {M_{\odot}}\else${M_{\odot}}$\fi}
\newcommand{\Lsun}{\ifmmode {L_{\odot}}\else${L_{\odot}}$\fi}
\newcommand{\Rsun}{\ifmmode {R_{\odot}}\else${R_{\odot}}$\fi}
\newcommand{\ergs}{erg~s$^{-1}$}
\newcommand{\ergcms}{erg~s$^{-1}$~cm$^{-2}$}
\newcommand{\radec}[9]{$\rmn{RA}(#1)=#2^{\rmn{h}}~#3^{\rmn{m}}~#4\fs#5$, $\rmn{Dec.}~(#1)=#6\degr~#7\arcmin~#8\farcs #9$}
\newcommand{\bepposax}{\textit{BeppoSAX}}
\newcommand{\swift}{\textit{Swift}}
\newcommand{\xrt}{\textit{Swift}/XRT}

\newcommand{\chandra}{\textit{Chandra}}
\newcommand{\acis}{\textit{Chandra}/ACIS}

\newcommand{\xmm}{\textit{XMM-Newton}}
\newcommand{\integral}{\textit{INTEGRAL}}
\newcommand{\rosat}{\textit{ROSAT}}
\newcommand{\srcigr}{IGR J17445$-$2747}
\newcommand{\srcsw}{Swift J175233.9$-$290952} 
\newcommand{\srcsxps}{1SXPS J174215.0$-$291453}
\newcommand{\srcxmm}{3XMM J174417.2$-$293944}
\newcommand{\sbsuvot}{Rivera~Sandoval et al., {\it in prep.}}

\graphicspath{{./}{figures/}}

\title[\swift\ Bulge Survey: X-ray Results]{The \swift\ Bulge Survey: Motivation, Strategy, and First X-ray Results}
\author[A. Bahramian et al.]{
A.~Bahramian,$^{1}$\thanks{E-mail: arash.bahramian@curtin.edu.au}
C.~O.~Heinke$^{2}$,
J.~A.~Kennea$^{3}$,
T.~J.~Maccarone$^{4}$,
P.~A.~Evans$^{5}$,
\newauthor
R.~Wijnands$^{6}$,
N.~Degenaar$^{6}$,
J.~J.~M.~in 't Zand$^{7}$,
A.~W.~Shaw$^{8,2}$,
L.~E.~Rivera~Sandoval$^{4,2}$,
\newauthor
S. McClure$^{2}$,
A. J. Tetarenko$^{9}$,
J.~Strader$^{10}$,
E.~Kuulkers$^{11}$,
G.~R.~Sivakoff$^{2}$
\\
$^{1}$International Centre for Radio Astronomy Research -- Curtin University, GPO Box U1987, Perth, WA 6845, Australia\\
$^{2}$Department of Physics, University of Alberta, CCIS 4-181, Edmonton, AB T6G 2E1, Canada\\
$^{3}$Department of Astronomy and Astrophysics, The Pennsylvania State University, University Park, PA 16802, USA\\
$^{4}$Department of Physics, Box 41051, Science Building, Texas Tech University, Lubbock, TX 79409-1051, USA\\
$^{5}$University of Leicester, X-ray and Observational Astronomy Group, School of Physics and Astronomy, University Road, Leicester, LE17RH, UK\\
$^{6}$Anton Pannekoek Institute for Astronomy, University of Amsterdam, Postbus 94249, NL-1090 GE Amsterdam, the Netherlands\\
$^{7}$SRON Netherlands Institute for Space Research, Sorbonnelaan 2, NL-3584 CA Utrecht, the Netherlands\\
$^{8}$Department of Physics, University of Nevada, Reno, NV 89557, USA\\
$^{9}$East Asian Observatory, 660 N. A'oh\={o}k\={u} Place, University Park, Hilo HI, USA, 96720\\
$^{10}$Center for Data Intensive and Time Domain Astronomy, Department of Physics and Astronomy, Michigan State University, East Lansing, MI 48824, USA\\
$^{11}$ESA/ESTEC, Keplerlaan 1, 2201, AZ Noordwijk, the Netherlands\\
}

\date{Accepted 2020 December 11. Received 2020 November 19; in original form 2020 September 18}

\pubyear{2020}

\begin{document}
\label{firstpage}
\pagerange{\pageref{firstpage}--\pageref{lastpage}}
\maketitle

\begin{abstract}
Very faint X-ray transients (VFXTs) are X-ray transients with peak X-ray luminosities ($L_X$) of L$_X\lsim10^{36}$ \ergs, which are not well-understood. We carried out a survey of 16 square degrees of the Galactic Bulge with the \swift\ Observatory, using short (60 s) exposures, and returning every 2 weeks for 19 epochs in 2017-18 (with a gap from November 2017 to February 2018, when the Bulge was in sun-constraint). Our main goal was to detect and study  VFXT behaviour in the Galactic Bulge across various classes of X-ray sources. In this work, we explain the observing strategy of the survey, compare our results with the expected number of source detections per class, and discuss the constraints from our survey on the Galactic VFXT population. We detected 91 X-ray sources, 25 of which have clearly varied by a factor of at least 10. 45 of these X-ray sources have known counterparts: 17 chromospherically active stars, 12 X-ray binaries, 5 cataclysmic variables (and 4 candidates), 3 symbiotic systems, 2 radio pulsars, 1 AGN, and a young star cluster. The other 46 are of previously undetermined nature. We utilize X-ray hardness ratios, searches for optical/infrared counterparts in published catalogs, and flux ratios from quiescence to outburst to constrain the nature of the unknown sources. Of these 46, 7 are newly discovered hard transients, which are likely VFXT X-ray binaries. Furthermore, we find strong new evidence for a symbiotic nature of 4 sources in our full sample, and new evidence for accretion power in 6 X-ray sources with optical counterparts. Our findings indicate that a large subset of VXFTs is likely made up of symbiotic systems. 
\end{abstract}



\section{Introduction}

Very Faint X-ray Transients (VFXTs) are phenomenologically defined as X-ray transients with peak X-ray luminosities in the range $10^{34}-10^{36}$ \ergs \citep{Wijnands06}.\footnote{Unless otherwise specified, $L_X$ and $F_X$ are given in 0.3--10 keV throughout this work.}
VFXTs have been relatively difficult to study, as they generally fall below the typical sensitivity of all-sky (or Galactic Bulge) monitors \citep[see surveys of][]{Swank01,intZand04,Kuulkers07,Krimm13,Negoro16}.
However, \chandra, \swift,  and \xmm\ surveys of the Galactic Centre have revealed that the majority of transient X-ray outbursts do not reach $10^{36}$ \ergs, and that VFXT outbursts may outnumber brighter X-ray transients \citep{Muno05, Degenaar09, Degenaar12b}.
Many of these transients have not shown any bright outbursts, though some objects show both very faint and brighter outbursts.

Various studies of VFXTs over the past decade have shown that these systems are an inhomogeneous population. A number of VFXTs have been identified as low-mass X-ray binaries (LMXBs; where compact objects accrete from donor stars of mass $<$2 \Msun) containing neutron stars \citep[e.g.,][]{Cornelisse02b, Degenaar12}.  These include some accreting millisecond X-ray pulsars (AMXPs, where the neutron star shows X-ray pulsations with millisecond periods), such as NGC 6440 X-2 \citep{Heinke10}, IGR J17062--6143 \citep{Strohmayer17}, and IGR J17379$-$3747 \citep{Sanna18}. Some VFXTs have been identified as LMXBs likely to contain black holes (e.g. CXOGC J174540.0$-$290031, \citealt{Bower05,Muno05b}; Swift J1357.2$-$0933, \citealt{Corral-Santana13, ArmasPadilla13b}; XTE J1728$-$295, \citealt{Sidoli11}).
Some VFXTs show slow X-ray pulsations that suggest a high-mass X-ray binary (HMXB) nature \citep[e.g.,][]{Torii98}. Low-luminosity HMXB behaviour has not been well-studied \citep[e.g.,][]{Wijnands16, RoucoEscorial17}.

Other objects, such as cataclysmic variables (CVs), magnetars, and symbiotic systems (in which a compact object accretes from a red giant wind) can also produce VFXT behaviour.   Novae may produce similar hard X-ray luminosities for a few weeks  \citep{Mukai08}, and the brightest intermediate polars (IPs) can reach this range \citep[e.g.][]{Brunschweiger09,Stacey11,Bernardini12,Suleimanov19}. Magnetars can also reach this X-ray luminosity range   \citep[e.g.,][]{CotiZelati18}, as can colliding-wind binaries \citep{Gudel09}, and background AGN will certainly be present.
In addition, a variety of foreground X-ray sources have been detected at similar fluxes in Galactic Centre surveys, thought to be dominated by nearby chromospherically active stars \citep[e.g.][]{Wijnands06}, but including other source populations such as cataclysmic variables. Since distances are not easy to determine, we often can not be certain whether detected transients fall into the VFXT luminosity range.

Some VFXTs have been detected in repeated outbursts, allowing assessment of their average mass-transfer rates. These inferred rates are quite low compared to normal (brighter) LMXBs, suggesting a later (or unusual) stage of LMXB binary evolution \citep[e.g.,][]{Degenaar09, Heinke15}. VFXT behaviour seems to be associated with parts of X-ray binary evolution that are not generally studied with brighter objects.  A likely candidate is the slow final evolution of $\sim$0.01 \Msun\ degenerate companions \citep{King06, intZand05, Heinke15}. LMXBs accreting from a weak donor wind before their Roche-lobe overflow phase, or while in the period gap, may also produce low-luminosity VFXT outbursts  \citep{Pfahl02c, Maccarone13}.  Black hole LMXBs with short orbital periods may also be extremely faint, if their radiative efficiency decreases with decreasing luminosity as predicted, and there should be large numbers of these short orbital period systems  \citep{Knevitt14,Maccarone15,Arur18}.

The evolution and final fate of LMXBs are difficult to model due to the wide range of physics that must be included. 
Major areas of uncertainty include the common envelope phase \citep[e.g.,][]{ivanova13}, effects of angular momentum losses \citep[e.g.,][]{Knigge11}, and the fraction of transferred mass accreted onto the compact object \citep[e.g.,][]{Ponti12}. Binary evolution calculations \citep[e.g.,][]{Patterson85} predict that the majority of LMXBs should have short orbital periods and low mass-transfer rates, and thus may have infrequent, short, and low-luminosity outbursts \citep{king00}. However, if angular momentum losses are much larger than in the standard evolution scenario \citep[e.g., due to strong winds from donors;][]{DiSalvo08,Marino19}, the lifetimes of LMXBs may be much shorter, and there may be many fewer LMXBs than predicted. Understanding the late stages of LMXB evolution (likely at low mass transfer rates, showing very faint outbursts) is critical for understanding how millisecond radio pulsars are born \citep[e.g.][]{Tauris12,Papitto14}.

We have begun a novel survey -  the \swift\ Galactic Bulge Survey (hereafter SBS), based on fast tiling procedures (see \S~\ref{sec:survey_config}) by the Neil Gehrels \swift\ Observatory \citep[hereafter \swift;][]{Gehrels04}, that uses biweekly shallow ($\sim$60-second) mosaic imaging by the X-ray telescope \citep[XRT;][]{Burrows05} and the Ultra-Violet/Optical Telescope \citep[UVOT;][]{Roming05} to identify VFXTs across 16 square degrees of the Galactic Bulge. Over the first year of this survey, we have detected multiple very faint outbursts and performed multi-wavelength follow-up observations to better understand the population and nature of VFXTs. 
This paper is one of three papers summarizing our findings from the first year of SBS (see also \citealt{Shaw20} and \sbsuvot, which detail results from follow-up observations with optical/near-IR facilities, and results from the \swift\ Ultraviolet/Optical Telescope (UVOT) data, respectively). In this paper, we report the plan of the survey and the initial X-ray results. We have also carried out a second year of SBS observations in 2019 and early 2020, which will be reported in detail elsewhere (some initial results are given in \citealt{Heinke19a,Heinke19b,Maccarone19,Heinke20b}).
We discuss the survey strategy and configuration in \S~\ref{sec:survey_config}, and the data reduction and analysis pipeline in \S~\ref{sec:analysis}. Finally we present the X-ray results and their implications in \S~\ref{sec:results}.

\section{The \swift\ Bulge Survey: strategy and configuration}\label{sec:survey_config}
To understand accretion at lower mass-transfer rates (and to understand the fraction of an X-ray binary's life spent in such states), we need surveys capable of identifying these relatively lower-luminosity outbursts. \swift's ability to tile large areas with very short XRT and UVOT observations meets this need. \swift's rapid slewing and quick operation permits exposures (with both XRT and UVOT) as short as 60 seconds, with 28 seconds between exposures for slewing and settling. \citet{Evans15} pioneered these new methods for the follow-up of gravitational wave and high-energy neutrino detections, and they have been used to detect moderately faint X-ray transients in a survey of the Small Magellanic Cloud \citep{Kennea18}.

The SBS consists of biweekly epochs, with an average of 120 pointings per epoch (up to 163) and $\sim$60 or 120 s per pointing. This has enabled regular repeated X-ray surveys of $\sim$16 square degrees around the Galactic Centre with a few arcsecond angular resolution, and sensitivity reaching to $\sim10^{35}$ \ergs\ at $\sim8$ kpc for $N_H\sim1\times10^{22}$ cm$^{-2}$. Since the Galactic Bulge is very rich in LMXBs, we estimate the distance of most LMXBs here to be $\sim$8 kpc. 

\begin{figure*}
\includegraphics[width=8cm]{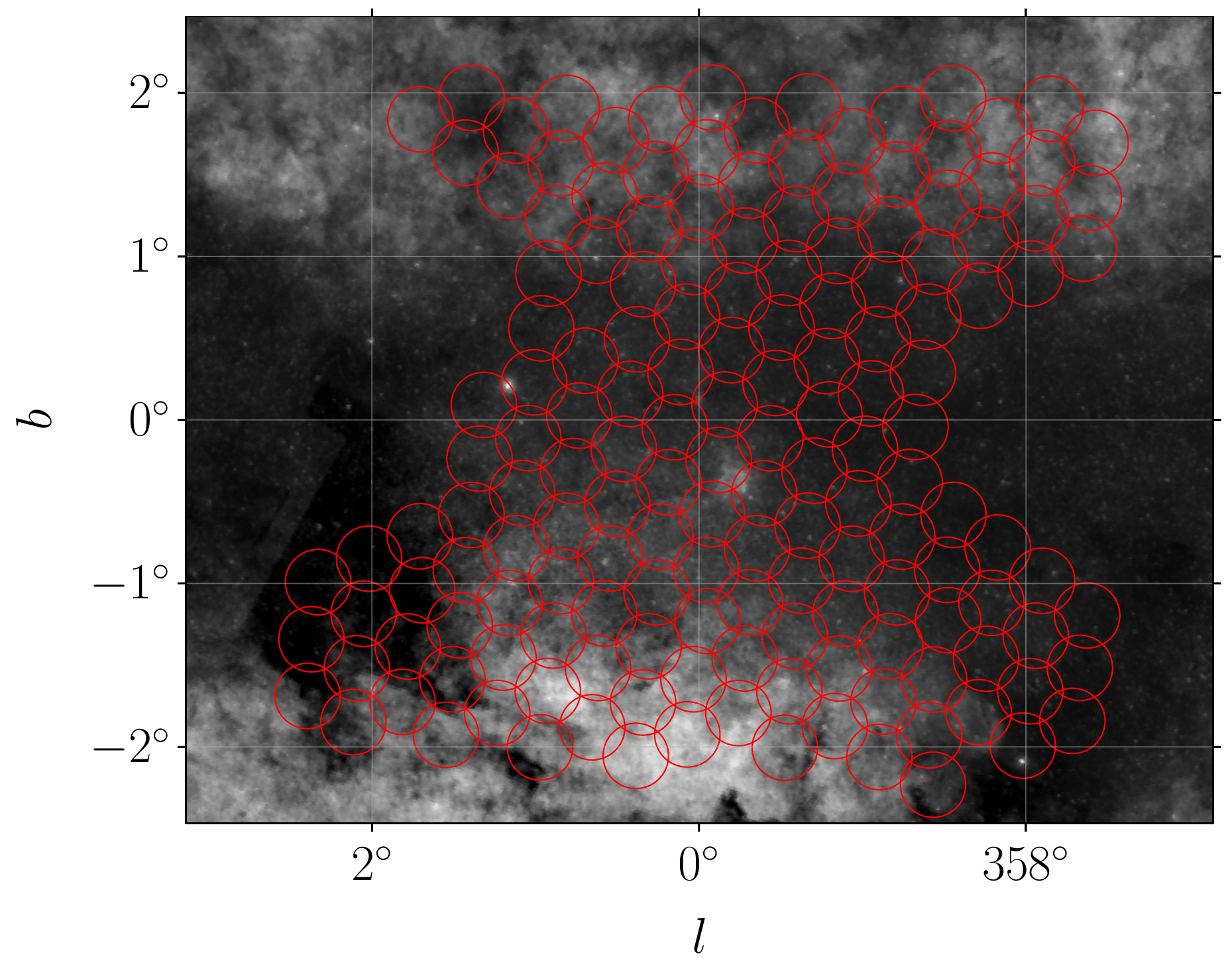}
\includegraphics[width=8.5cm]{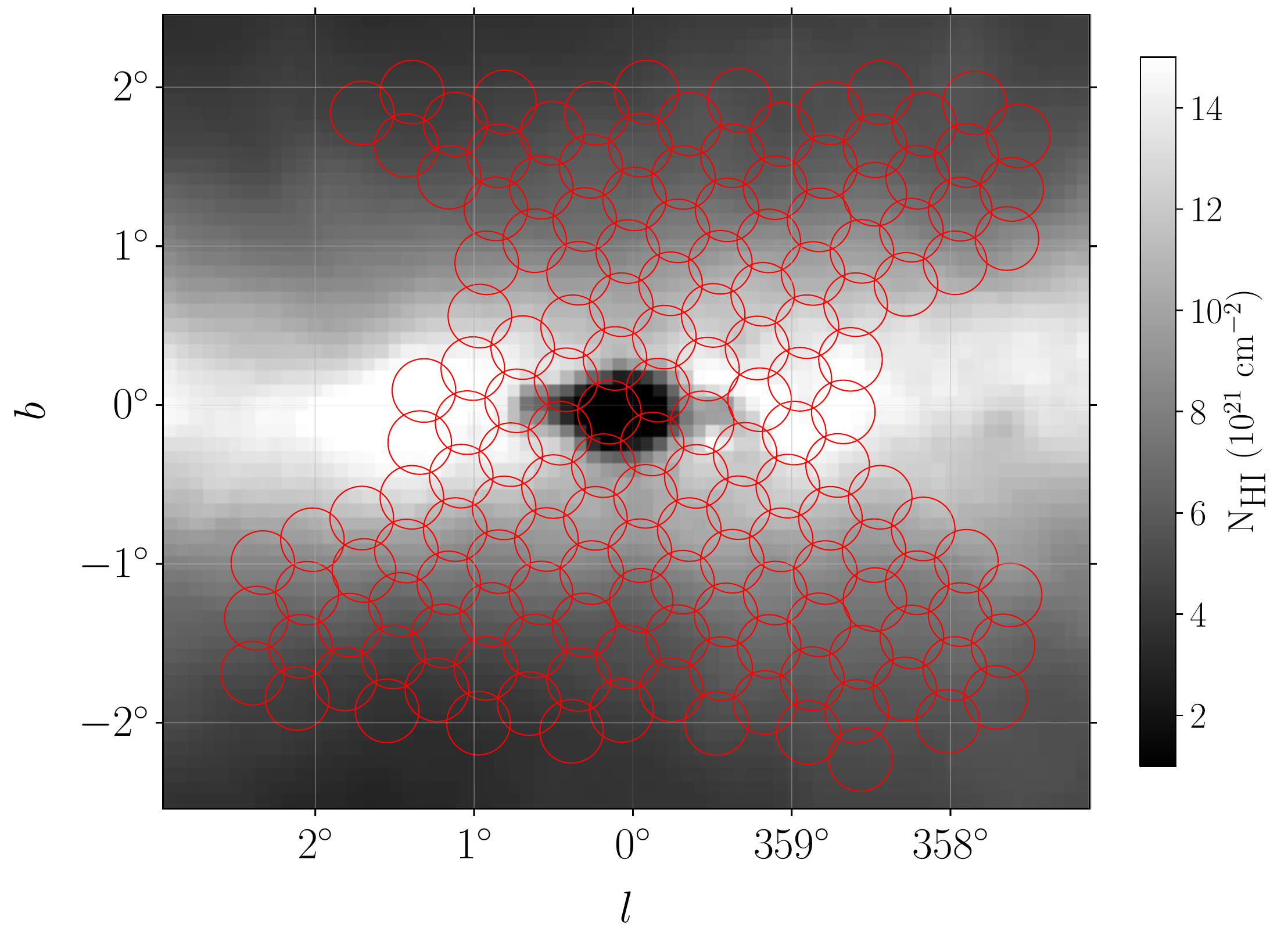}
\caption{\swift\ Bulge Survey coverage footprint during the first year of the survey (in epochs 2 to 15, after exclusion of the area affected by the stray light from GX 3+1, see text and Figure~\ref{fig:sbsimage}), superposed on the image of the Galactic Bulge (left; from the STScI Digitized Sky Survey) and map of neutral hydrogen column density (right; from \citealt{HI4PI2016}). The survey shape was chosen as a compromise between the concentration of LMXBs towards the Galactic Centre, and the increased extinction towards the Plane. The dark patch at the centre in the HI map is caused by the absence of accurate measurements, and is not real.}
\label{fig:pointings}
\end{figure*}

The size and shape of the survey area was selected as a compromise between the concentration of LMXBs towards the Galactic Centre, and the increased extinction towards the Galactic plane (Fig.~\ref{fig:pointings}). For example, at 1--2 degrees off the Galactic plane, the hydrogen column density ($N_H$) falls from $\gsim10^{23}$ cm$^{-2}$ to $\lsim 10^{22}$ cm$^{-2}$, allowing detection of activity down to  $L_X \sim 4\times10^{34}$ \ergs. The aim of the SBS is to ensure a large number of VFXT detections (thus enclosing regions around the Galactic Centre), while also covering a larger region close to the Galactic plane with relatively low extinction and good multiwavelength coverage (e.g., the \chandra\ Galactic Bulge Survey fields, located 1--2 degrees above and below the Galactic plane; \citealt{Jonker11}).

\begin{figure*}
\includegraphics[width=14cm]{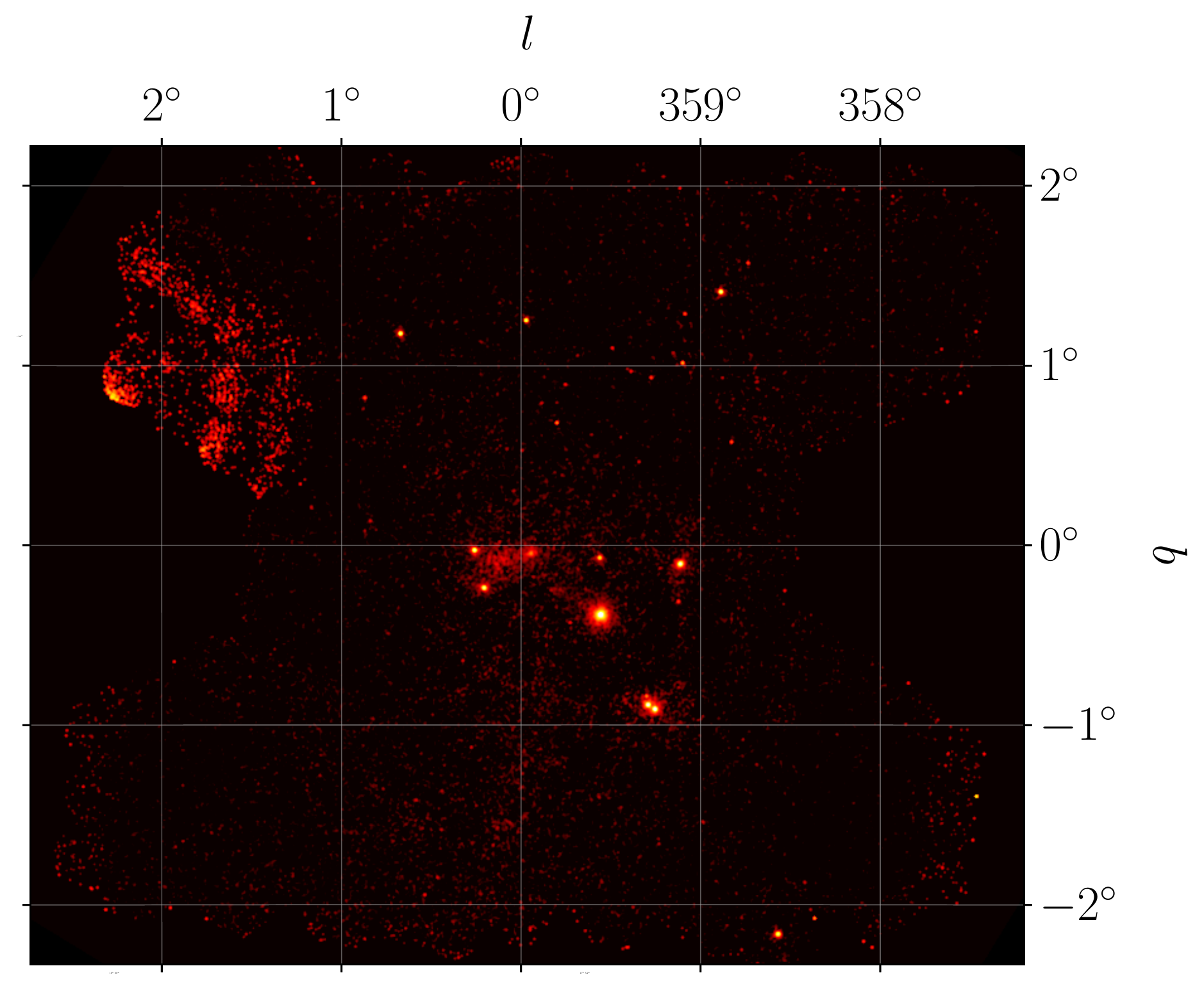}
\includegraphics[width=14cm]{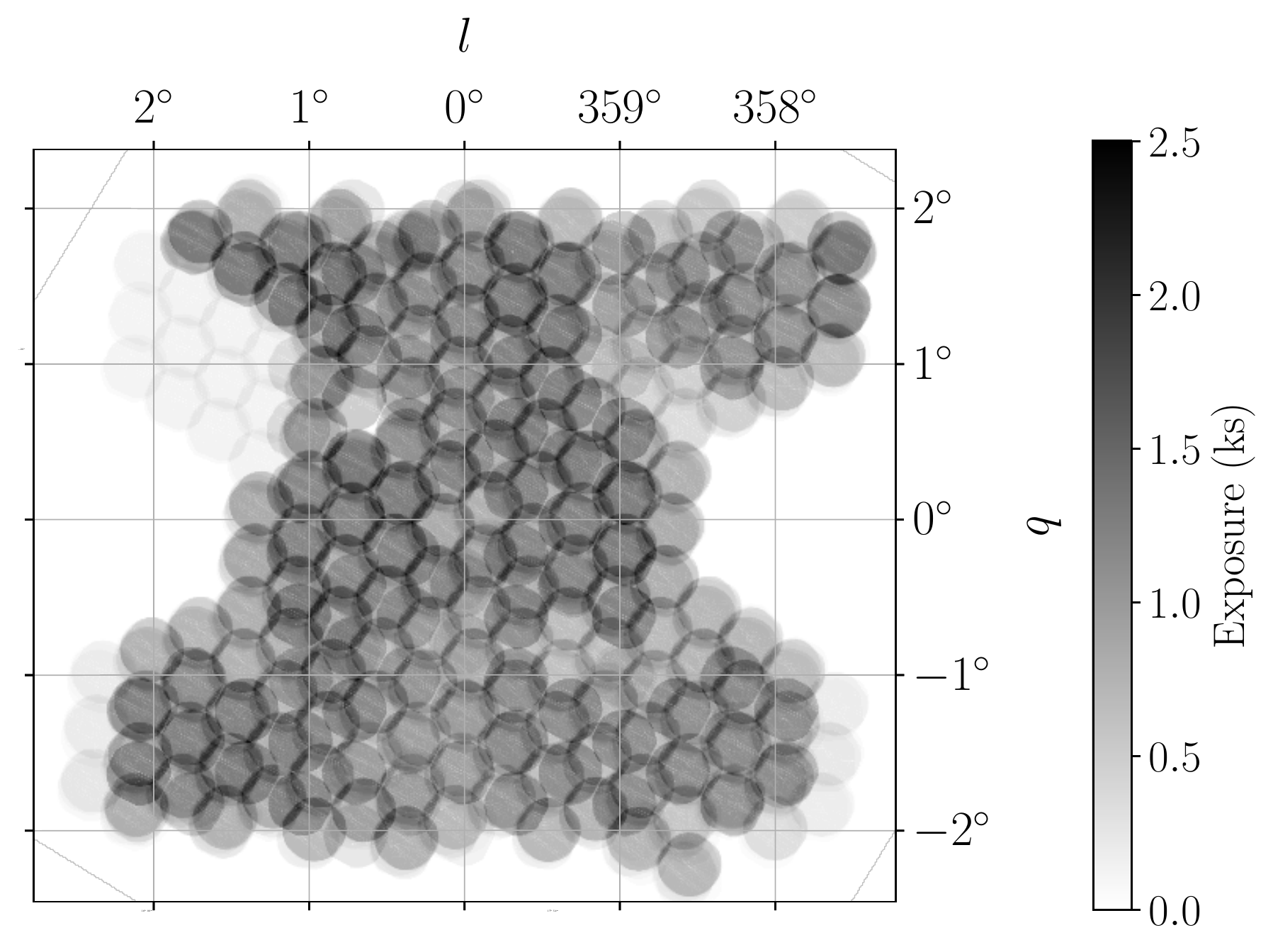}
\caption{Top - \xrt\ stacked and exposure-corrected mosaic image of the \swift\ bulge survey in the first year. The north-east corner of the field (upper left of the image) is heavily contaminated by scattered light from the LMXB GX 3+1. This region was omitted after the first epoch. Bottom - The stacked \xrt\ exposure map mosaic of the first year of SBS.} 
\label{fig:sbsimage}
\end{figure*}

In the first year of this survey (April 2017 -- March 2018), we obtained a total of 19 biweekly epochs with an average XRT exposure time of 50-60 seconds per pointing for epochs 1 to 15 and 115-120 seconds for epochs 16-19. Our survey strategy and pointing pattern was slightly adjusted through the course of the survey based on previous epochs. For example, after the first epoch we noticed that scattered light from the bright persistent LMXB GX 3+1 substantially contaminates some of the survey tiles located $\lesssim$ 1$^{\circ}$ away from GX 3+1 (Fig.~\ref{fig:sbsimage}, see \citealt{Evans20} for a detailed discussion and modeling of this artifact in \xrt). Thus we excluded these tiles in later epochs. In the last 4 epochs, we reduced the number of tiles while doubling individual exposures, to increase depth of individual epochs, and reduced the overlap area between tiles to retain a large coverage area (see Section 4.6). Details of the survey epochs are summarized in Table~\ref{tab:obslist}.

\begin{table}
\centering
\caption{Summary of \swift\ Bulge Survey epochs and observations. The 4-month gap between epochs 15 and 16 was caused by \swift\ satellite pointing Sun constraints in the direction of the Galactic Bulge. Epochs 16-19 were performed with the redesigned strategy (fewer, more widely-spaced number of exposures with higher exposure times per individual snapshots). $a$- Epoch 6 was performed over two segments a week apart. $b$- Epoch 15 was interrupted by a high-priority transient event.}
\label{tab:obslist}
\begin{tabular}{lcccc}
\hline
\hline
Epoch   &   Start time          & N Tiles & Tile Exp. & Total Exp. \\
        &   (UTC)               &         & (s)           & (ks)            \\
\hline
1       &   2017-04-13 00:21:12 & 139     & 58            &   8.1           \\
2       &   2017-04-20 04:32:57 & 148     & 58            &   8.6           \\
3       &   2017-05-04 00:11:07 & 144     & 58            &   8.3           \\
4       &   2017-05-18 04:14:01 & 145     & 55            &   7.9           \\
5       &   2017-06-01 02:49:22 & 128     & 49            &   6.3           \\
6$^a$   &   2017-06-15 03:01:01 & 153     & 47            &   7.2           \\
7       &   2017-06-29 00:11:59 & 90      & 50            &   4.5           \\
8       &   2017-07-13 08:46:24 & 121     & 58            &   7.1           \\
9       &   2017-07-27 06:11:43 & 163     & 59            &   9.6           \\
10      &   2017-08-10 00:02:35 & 133     & 58            &   7.7           \\
11      &   2017-08-24 05:13:44 & 150     & 57            &   8.6           \\
12      &   2017-09-07 03:50:35 & 109     & 58            &   6.6           \\
13      &   2017-09-21 02:45:55 & 130     & 58            &   7.6           \\
14      &   2017-10-05 00:02:36 & 158     & 57            &   9.2           \\
15$^b$  &   2017-10-19 01:56:24 & 61      & 57            &   3.5           \\
16      &   2018-02-15 05:40:57 & 93      & 119           &   11.1          \\
17      &   2018-02-28 23:58:57 & 93      & 116           &   10.8          \\
18      &   2018-03-15 07:50:57 & 65      & 117           &   7.6           \\
19      &   2018-03-29 00:18:56 & 86      & 116           &   10.0          \\
\hline
\end{tabular}
\end{table}

\section{\xrt\ Data reduction and analysis}\label{sec:analysis}
\subsection{Reduction and analysis pipelines}
The primary goal of this survey is a search for VXFTs, many of which show short ($\sim$ hours--days) transient activity. This requires reduction and analysis of the survey data as soon as possible (e.g., by obtaining preliminary data from the Quick-look database). However, Quick-look data are not the final processing of the data, and can be updated with further telemetry or processing. Thus we developed two pipelines, one for rapid processing of data as soon as they become available on the \swift\ Quick-look database\footnote{\url{https://www.swift.ac.uk/archive/ql.php}}, and another pipeline with more robust calibration and more careful processing, but with substantially higher overhead, for analysis of final archival data in the \swift\ database\footnote{\url{https://www.swift.ac.uk/archive/obs.php}}.

Both pipelines follow fundamentally similar procedures: 1- identification and acquisition of observations that belong to a specific epoch in the survey (as tabulated in Table~\ref{tab:obslist}); 2- performing event energy filtering and creating image files from events; 3- producing mosaic images and exposure maps for the epoch; 4- performing source detection and producing a source catalogue for each epoch; 5- cross-matching detections against existing catalogs (e.g., via Vizier and SIMBAD databases). 

The only difference between the the two pipelines is the status and reduction of the initially acquired data sets. The focus of the Quick-look pipeline is rapid analysis (as opposed to robust calibration and reduction). Thus, the Quick-look pipeline obtains a pre-processed ``clean'' event file and exposure map for each observation from the \swift\ archive, as opposed to locally produced versions of those files. Some of these clean event files were produced using older versions of \textsc{heasoft}\footnote{\url{https://heasarc.gsfc.nasa.gov/docs/software/heasoft}} (versions $<6.22$). Small version differences typically do not cause significant discrepancies. However, the newest updates to the \texttt{xrtpipeline}, introduced in 2017 (version 0.13.4, included in \textsc{heasoft} 6.22 and higher)\footnote{\url{https://heasarc.gsfc.nasa.gov/docs/software/lheasoft/Rel_Notes_6.22.html}}, impact exposure filtering in short \xrt\ observations significantly. In contrast, using the robust pipeline, we reprocess level 1 \xrt\ data of archived observations locally, using \textsc{heasoft} 6.25. This procedure introduces a significant overhead compared to the Quick-look pipeline, but allows a more careful examination of the data (without loss of events due to exposure filtering, or incomplete telemetry). The rest of the reduction and analysis procedure is identical between the two pipelines.

We then produce filtered event and image files for each observation in the 0.5--10, 1.5--10, and 0.5--1.5 keV bands using \texttt{xselect}, and make wide-field mosaic images and exposure maps for each epoch. For producing mosaic images and exposure maps, we used the \textsc{Swarp} package\footnote{\url{http://astromatic.iap.fr/software/swarp}} \citep[version 2.38,][]{Bertin02} with the combination method set to \texttt{sum} and background subtraction disabled. 

The 0.5--1.5 keV and 1.5--10 keV bands were chosen specifically to estimate source hardness, and possibly to constrain distance. Due to the extremely high extinction in the direction of the Galactic Bulge, these energy bands can be used as a very rough proxy for distance. Soft ($\lsim1.5$ keV) X-ray emission from sources at $d\gsim5$ kpc in this region is likely to be mostly extinguished by interstellar absorption \citep[e.g., see][]{Schultheis14}. 
As an example, we calculate (using PIMMS\footnote{\url{https://asc.harvard.edu/toolkit/pimms.jsp}}) that in the lowest-$N_H$ portion of our field ($l$=0, $b$=-2 deg), where $N_H\sim7\times10^{21}$ cm$^{-2}$, a background X-ray source with photon index 1.7 will show a hardness ratio (1.5--10 keV)/(0.3--10 keV) of 0.8, while an unabsorbed source of the same index will have a hardness ratio of 0.5. 
Thus we expect that hardness ratios $\gsim0.6$ indicate substantial absorption, while lower values suggest foreground objects.

\subsection{Rapid Source Detection}\label{sec:detection}
The low exposure in individual snapshots combined with the rather large size of the \xrt\ point spread function \citep[18$''$ half-power diameter;][]{Moretti05} make source detection challenging in the photon-starved regime dominating our survey. Detecting faint sources (with \xrt\ count rate $\leq 0.5$ ct s$^{-1}$) in 60 s snapshots is especially difficult and unreliable, due to low significance and high false detection rates.

Given these issues, we developed the following source detection method for detecting transients in each epoch: We first convolve the mosaic image of an epoch with a Gaussian kernel with a standard deviation of 10 pixels using the \texttt{convolution} package in \textsc{Astropy} \citep{Robitaille13}. Then, we use the task \texttt{find\_peaks} in the package \textsc{Photutils}\footnote{\url{https://photutils.readthedocs.io/}} \citep{Bradley2019} to perform source detection on the smoothed image. We performed two sets of detections, with different detection thresholds, representing high ($>$ 3-$\sigma$) and low (1- to  3-$\sigma$) significance levels based on the average background in the images. Then, we checked for detections in all previous \xrt\ coverage of the area, and against other X-ray (and multi-wavelength) catalogs. 

This method allowed us to identify faint sources effectively across our survey. A substantial fraction of the false detections picked up by this detection method could be ruled out by visual inspection based on their shape/location. For example, some of these detections were located in the contaminated regions around GX 3+1, or within the diffuse emission region near the Galactic Centre (Fig.~\ref{fig:sbsimage}, left), or near other bright sources. Another group were caused by optical leak of photons from extremely bright optical/ultraviolet stars (e.g., from V* X Sgr). We subjected all high-significance detections of new sources to X-ray, and often optical and/or near-IR, follow-up, typically within 1-7 days (see \citealt{Shaw20} for details on some of these observations and their conclusions).

\subsection{\swift\ Bulge Survey X-ray Source Catalogue}\label{sec:catalog}

To construct a final source catalogue we used an approach almost identical to that employed for the \swift\ Small Magellanic Cloud Survey \citep[S-CUBED, ][]{Kennea18}, except that we based it on the new XRT catalogue tools created for the \swift-XRT Point Source catalogue \citep[2SXPS, ][]{Evans20}. In this approach, the XRT fields were grouped into `blocks' of overlapping fields, with each block being up to $\sim0.6\deg$ in radius. Blocks were defined such that every field, and every overlap between fields, was in at least one block. Source detection, localisation and characterisation were then carried out using the 2SXPS software \citep{Evans20}, which produces positions and fluxes for every source, as well as a flag characterising the probability that it is spurious. These flags have values of `Good', `Reasonable' and `Poor', with false positive rates of 0.3\% for good sources, 1\% for the set of good+reasonable sources and 10\% when poor sources are also included. Due to the large amount of diffuse emission around the Galactic Bulge, especially around the Galactic Centre, the rate of spurious sources is expected to be somewhat elevated\footnote{In 2SXPS extra warnings are set in the flags in such cases; these warnings were not set in our analysis.}, and in particular we found that a very high fraction of `poor' sources were clearly spurious. We therefore limit the results reported here to `good' and `reasonable' sources.

This pipeline produced a total of 831 individual sources, with varying level of significance. Out of these, 654 detections are of poor quality, 54 of reasonable quality, and 123 were good detections. For a more careful investigation, we omitted poor-quality detections, and inspected each of the remaining 177 individually. We found that a small subset of these detections are from areas suspected of diffuse emission (e.g., around Sgr A*), in the PSF of very bright sources (e.g., around H 1742$-$294), or likely due to optical loading in the vicinity of very bright stars (e.g., around V* X Sgr). Ignoring these detections, along with other instances where a visual inspection identifies an unreliable detection, resulted in a final catalogue of 104 sources (shown in part in Tables~\ref{tab:srccat_good}, \ref{tab:srccat_spec}, \ref{tab:srccat_counterpart}, \ref{tab:srccat_nature}; the complete catalogue is available in the electronic version of the manuscript). Many of these sources are previously identified X-ray emitters. However, for most of them, the SBS provides the first long-term high-cadence variability study, allowing detection of faint outbursts, among other long-term behaviors. We crossmatched the sources in this catalogue against SIMBAD \citep{Wenger00}, the \chandra\ Source Catalog\footnote{\url{https://cxc.harvard.edu/csc/}} \citep{Evans10} release 2.0 (2019), and the \xmm\ source catalogue 3XMM-DR8 \footnote{\url{http://xmmssc.irap.omp.eu/Catalogue/3XMM-DR8/3XMM_DR8.html}} \citep{Rosen16}.
To estimate the number of possible chance coincident matches, we produced multiple sets of mock localisations by shifting each SBS source by $\sim 1$ arcmin in an arbitrary direction (to imitate a randomly generated source catalogue with similar global source distribution as the SBS catalogue). These randomisations typically produced $\leq 4$ matches between the catalogues, limiting the expected number of coincident matches.

Although we measure variations between our Swift observations, these are not sensitive to variability for most of our sources, which are too faint for a nondetection to be statistically meaningful, and thus we cannot robustly determine whether most of our sources are transients using Swift data alone. Luckily, we often have the opportunity to detect transient behaviour by comparing to archival \chandra\ or \xmm\ observations, either detections or nondetections. 
The fields of the majority of our SBS transients have been imaged with \chandra\ (especially by the Galactic Bulge Survey, \citealt{Jonker11,Jonker14}, which gives positional names starting with CXOGBS, and shorthand CX names) and/or \xmm. This allows us to use the ratio of the peak flux in the SBS and the (likely) quiescent \chandra\ or \xmm\ flux, based on the catalogued values (from the catalogues above), or nondetections, to indicate transient behaviour. 
 Typical nondetections come from 2 ks \chandra\ Galactic Bulge Survey \acis\ observations, for which we used an upper limit of $F_X$(0.3--10 keV, absorbed)$=8\times10^{-14}$ \ergcms (assuming a conservative 5 count detection limit, a high $N_H=6\times10^{22}$ cm$^{-2}$ - based on spectral analysis of sources in the bulge from our survey, and a power-law of photon index 1.7).
 We classify a source as transient if the peak SBS flux lower limit is $>$10 times the quiescent flux, and as a likely transient if the peak SBS flux lower limit is at least 5 times the quiescent flux (as our SBS errors are often large). We find 25 sources that have clearly varied by at least a factor of 10, and 14 more that have varied by at least a factor of 5; this is a lower limit on the variability among our X-ray sources, as many of our Swift detections are highly uncertain.

With this process, we were able to determine possible classifications for 45 sources in the catalogue, which we discuss in the following sections.

\begin{figure*}
\includegraphics[width=16cm]{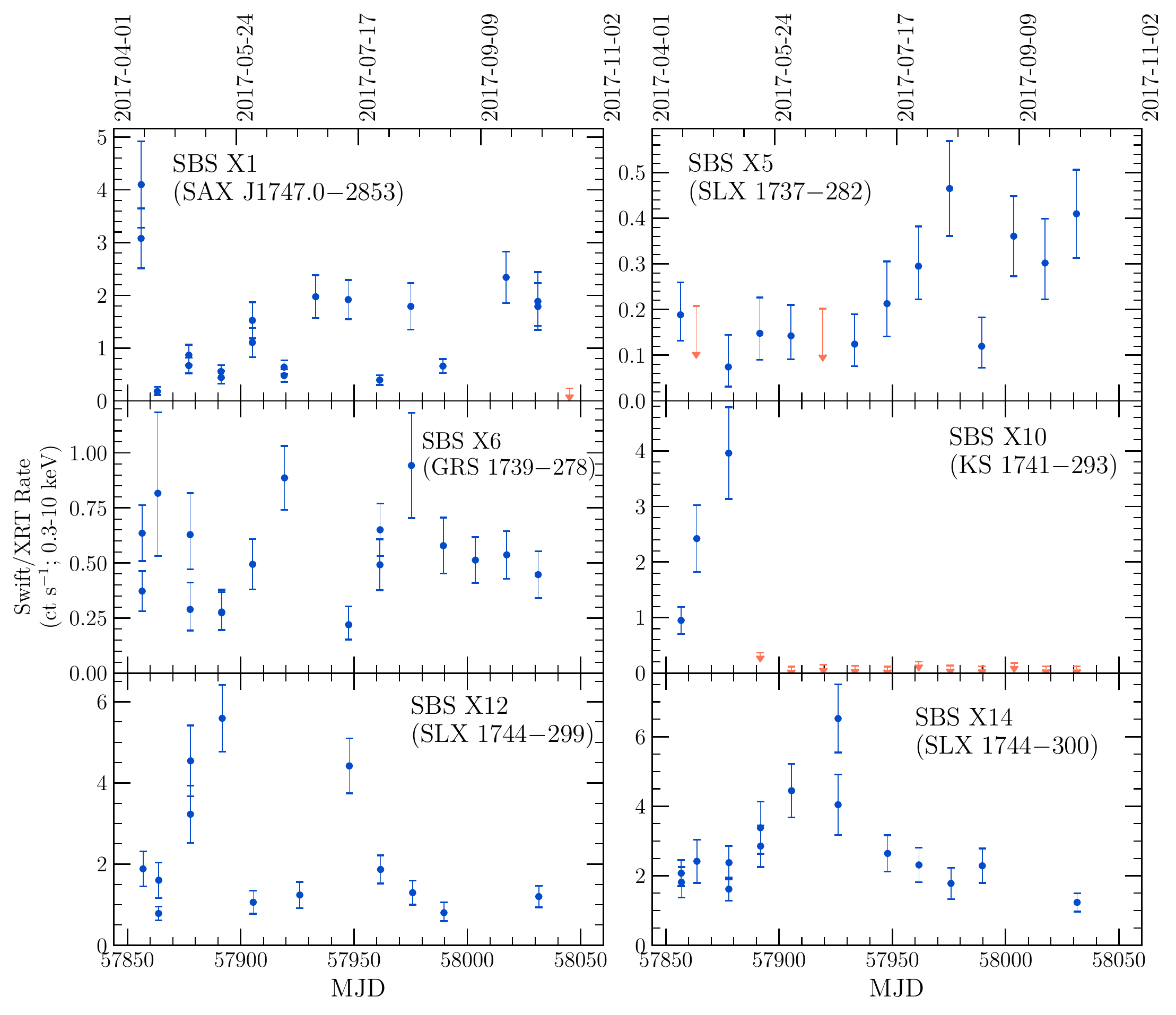}
\caption{\xrt\ light curves for a handful of sources (LMXBs in this example, see \S\ref{sec:xrbs}) in the SBS. The complete figure-set (containing light curves for all SBS sources) is available in the electronic version of this publication.}
\label{fig:lc_sample}
\end{figure*}

Tables~\ref{tab:srccat_good}, \ref{tab:srccat_spec}, \ref{tab:srccat_counterpart} and \ref{tab:srccat_nature} represent catalogue columns for a subset of sources detected in our survey. These include source designations, detection quality, coordinates, possible counterparts, basic estimates of peak count rate and absorbed flux (in the 0.3--10 keV band), hardness ratio (fraction of flux in 1.5--10 keV), results of X-ray spectral analysis, and a simple measure of variability significance (Var-$\sigma$) defined as the difference between peak count rate and the lowest detection (or upper limit) within the SBS data, divided by the 1-sigma uncertainty. For each source, we fitted the X-ray spectrum (stacked across all observations in which the source was detected) with an absorbed power-law using Bayesian inference and nested sampling using \textsc{Xspec}, and \textsc{Bxa} \citep{Arnaud96, Buchner14, Feroz19}. Table \ref{tab:srccat_counterpart} contains flux measurements of likely counterparts for each SBS source in the \chandra\ and \xmm\ source catalogues. We also provide a flux ratio between the measurements from those catalogues and the peak value from the SBS catalogue (converted to the appropriate band for each instrument, using the \textsc{pimms}\footnote{\url{https://asc.harvard.edu/toolkit/pimms.jsp}} software and assuming a power-law fit with parameters based on the merged spectrum in SBS, Table~\ref{tab:srccat_spec}). We also produce light curve plots for all the sources in the catalogue. These figures are available in the online version of the journal, and a subset are presented in Fig.~\ref{fig:lc_sample}.

\begin{figure*}
\includegraphics[width=8.5cm]{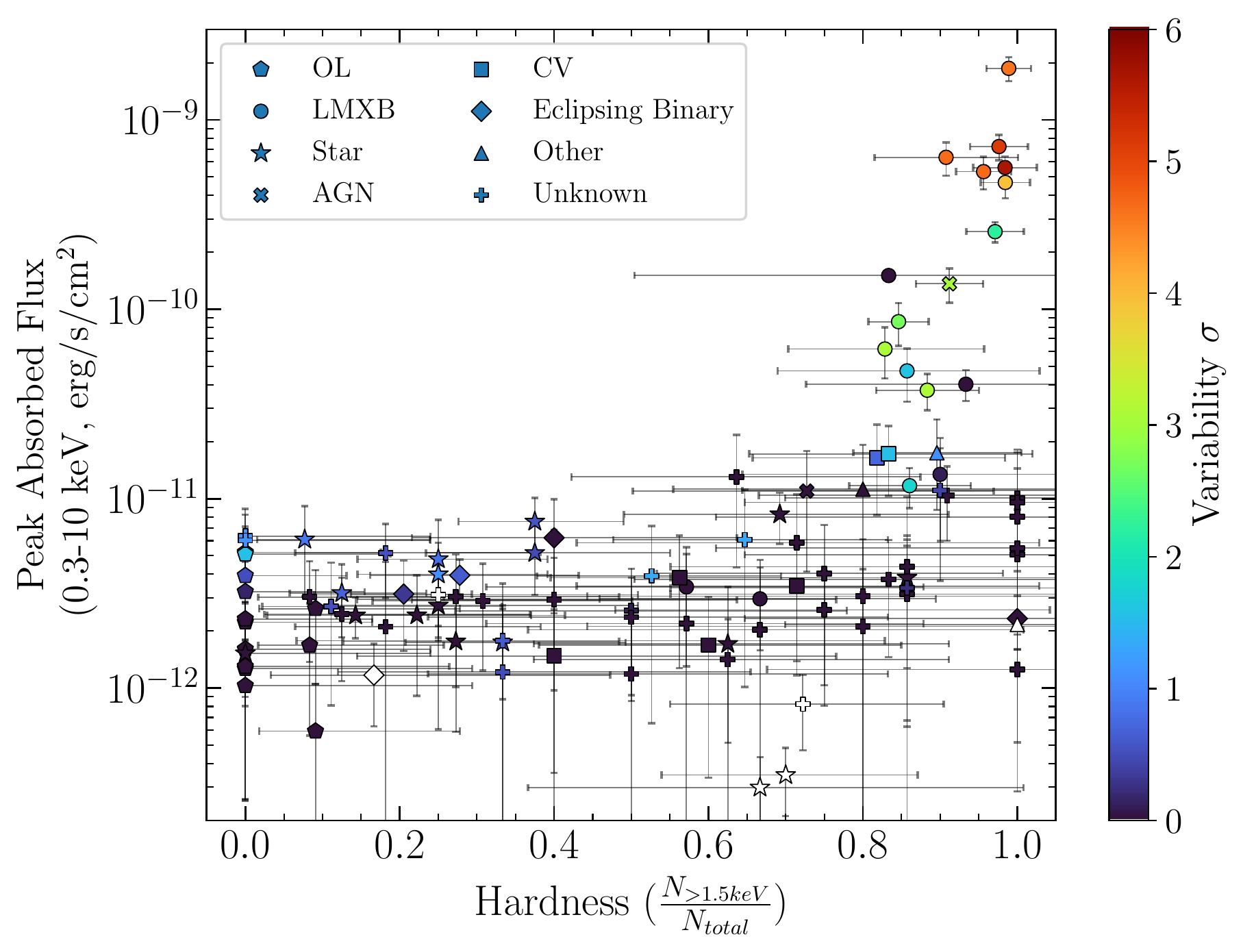}
\includegraphics[width=8.9cm]{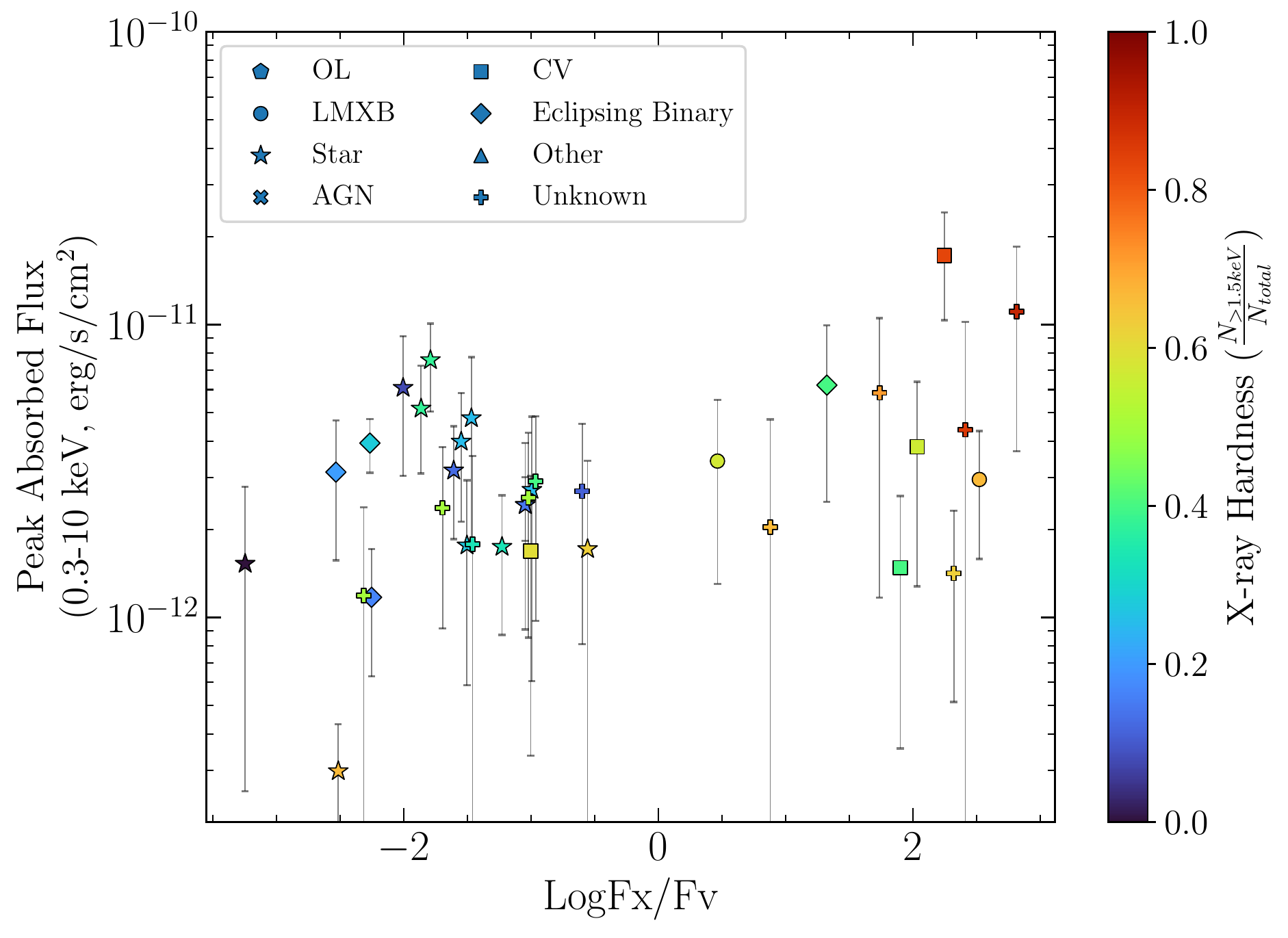}
\caption{{\bf Left}: Flux, hardness ratio (fraction of counts in 1.5--10 keV), and variability within SBS of sources in the SBS catalog. Harder sources are likely located in the Galactic Bulge (or beyond), while soft sources are likely in the foreground. ``OL'' stands for detections caused by the optical loading artifact (see \S\ref{sec:ol}). White markers are detections for which no variability properties could be inferred (e.g., detections in the stacked image). Variability significance is defined as the difference between peak count rate and the lowest detection (or upper limit) within the SBS data, divided by the 1-sigma uncertainty. {\bf Right}: X-ray flux versus X-ray-to-optical flux ratio for sources with clear counterparts.} 
\label{fig:sbscatalog}
\end{figure*}

The combination of observed flux, hardness ratio, and variability measure provide a simple picture of detected sources in the first year of our survey (Fig.~\ref{fig:sbscatalog}).

We identify emission from 17 bright chromospherically active stars, 12 X-ray binaries,  5 CVs, 3 symbiotic systems, 1 AGN, 2 radio pulsars, 1 young massive cluster, and 4 candidate CVs.  We discuss their identifications and properties below, and then discuss systems whose nature is not yet known.

\begin{table*}
\centering
\caption{SBS Catalog Part I: Observed properties of sources in the SBS. Position uncertainty (Err) corresponds to the 90\%-confidence \xrt\ radial uncertainty. The peak count rate is reported in the 0.3--10 keV band. \#Detect/Cover indicates number of snapshots in which the source was detected versus the total number of snapshots covering the source region. 0 detections means the source was only detected in the stacked image, as opposed to any single snapshot. Var-$\sigma$ represents variability confidence based on peak rate and the lowest detection (or upper limit) within the SBS data divided by the 1-sigma uncertainty. The complete catalog is available in the electronic version of this publication.}
\label{tab:srccat_good}
\begin{tabular}{cccccccccc}
    \hline
    \hline
SBS X	& \swift\	                & Quality   	& RA	    & Dec	        & Err	        & Peak Rate		    & Hardness	        & \#Detect/Cover& Var-$\sigma$\\
        &                       &               & (\degr)   & (\degr)       & ($''$)        & (ct s$^{-1}$)     &                   &               &           \\
\hline
1	    & J174702.6$-$285259	& Good	        & 266.76118	& -28.8833	& $3\farcs5$	& $4.1\pm0.8$       & $1.0\pm0.1$         & 19/20	        & 4.7        \\
2	    & J174621.1$-$284342	& Good	        & 266.58812	& -28.7284	& $3\farcs5$	& $1.2\pm0.3$       & $1.0\pm0.1$         & 15/15	        & 2.2        \\
3	    & J174627.1$-$271122	& Reasonable	& 266.61314	& -27.1897	& $8\farcs8$	& $0.04\pm0.02$     & $0.2_{-0.1}^{+0.3}$ & 0/5	            & --         \\
4	    & J174430.5$-$274600	& Good	        & 266.12713	& -27.7667	& $5\farcs4$	& $0.5\pm0.2$       & $0.8\pm0.1$         & 4/14	        & 3.1        \\
5	    & J174042.8$-$281807	& Good	        & 265.17861	& -28.3020	& $3\farcs7$	& $0.5\pm0.1$       & $0.9\pm0.1$         & 12/14	        & 3.1        \\
6	    & J174240.0$-$274455	& Good	        & 265.66678	& -27.7487	& $3\farcs9$	& $0.9\pm0.2$       & $0.8\pm0.1$         & 17/17	        & 2.7        \\
7	    & J174230.3$-$284454	& Good	        & 265.62663	& -28.7484	& $4\farcs2$	& $0.12\pm0.06$     & $0.2\pm0.1$         & 12/13	        & 0.3        \\
8	    & J174354.8$-$294441	& Good	        & 265.97867	& -29.7450	& $3\farcs5$	& $2.3\pm0.4$       & $1.0\pm0.1$         & 16/16	        & 3.9        \\
9	    & J174445.2$-$295045	& Good	        & 266.18842	& -29.8459	& $4\farcs3$	& $0.3\pm0.1$       & $0.9\pm0.2$         & 4/16	        & 1.6        \\
10	    & J174451.5$-$292042	& Good	        & 266.21483	& -29.3451	& $3\farcs9$	& $4.0\pm0.8$       & $0.9\pm0.1$         & 3/14	        & 4.7        \\
\hline
\end{tabular}
\end{table*}

\begin{table*}
\centering
\caption{SBS Catalog Part II: Spectral analysis. All spectra were fit with an absorbed power-law. Flux values are reported in the 0.3--10 keV band (median of the posterior distribution), and uncertainties indicate the 5\% and 95\% credible interval on posterior distributions. Peak fluxes are estimated based on the peak count rates, assuming the spectral model from the stacked spectra, converted using PIMMS. The complete catalog is available in the electronic version of this publication.}
\label{tab:srccat_spec}
\begin{tabular}{ccccccc}
    \hline
    \hline
SBS X	& \swift\ 	                &   N$_H$   	                    &   $\Gamma$            &   Avg. Unabs. Flx             &   Avg. Abs. Flx               &   Peak Abs. Flx           \\
        &                       &   ($10^{21}$ cm$^{-2}$)           &                       &   ($10^{-12}$ \ergcms)        &   ($10^{-12}$ \ergcms)        &   ($10^{-12}$ \ergcms)    \\
\hline
1	    & J174702.6$-$285259    &	$134_{-25}^{+31}$	            &   $2.1\pm0.5$         &   $344_{-137}^{+423}$ 	    &   $73_{-30}^{+91}$	    & $533\pm104$   \\
2	    & J174621.1$-$284342    &	$249_{-55}^{+62}$	            &   $1.1_{-0.5}^{+0.6}$ &   $245_{-51}^{+139}$	        &   $104_{-22}^{+59}$	    & $257\pm32$    \\
3	    & J174627.1$-$271122    &	$0.74_{-0.71}^{+3.8}$	        &   $2.6\pm1.1$         &   $1.7_{-1.0}^{+3.0}$	        &   $1.3_{-0.7}^{+2.2}$     & $1.2\pm0.5$       \\
4	    & J174430.5$-$274600    &	$3.23_{-3.19}^{+45.20}$	        &   $0.3_{-0.6}^{+1.6}$ &   $4.9_{-1.5}^{+3.5}$	        &   $4.8_{-1.5}^{+3.4}$	    & $62\pm19$     \\
5	    & J174042.8$-$281807    &	$27\pm7$                	    &   $1.9\pm0.4$         &   $39_{-9}^{+19}$ 	        &   $18.1_{-4.1}^{+8.8}$    & $37\pm8$      \\
6	    & J174240.0$-$274455    &	$25_{-5}^{+6}$	                &   $1.5\pm0.3$         &   $59_{-6}^{+10}$ 	        &   $36_{-4}^{+6}$  	    & $86\pm22$     \\
7	    & J174230.3$-$284454    &	$3.5_{-1.7}^{+1.6}$	            &   $3.4_{-0.7}^{+0.5}$ &   $6_{-3}^{+5}$	            &   $1.3_{-0.7}^{+1.2}$	    & $3.1\pm1.6$       \\
8	    & J174354.8$-$294441    &	$150_{-30}^{+35}$	            &   $0.6\pm0.4$         &   $420_{-43}^{+67}$	        &   $261_{-27}^{+42}$	    & $467\pm81$    \\
9	    & J174445.2$-$295045    &	$218_{-136}^{+165}$	            &   $2.4_{-2.0}^{+1.3}$ &   $27_{-21}^{+422}$	        &   $3.4_{-2.6}^{+52}$	    & $47\pm15$     \\
10	    & J174451.5$-$292042    &	$378_{-128}^{+110}$	            &   $3.1_{-1.2}^{+0.7}$ &   $1200_{-1100}^{+4800}$  	&   $24_{-21}^{+94}$	    & $634\pm127$   \\
\hline
\end{tabular}
\end{table*}

\begin{table*}
\centering
\caption{SBS Catalog Part III: catalogue columns containing information on previous detections in the X-rays by \chandra\ and \xmm, and magnitude of the possible optical/infrared counterpart. All fluxes are in units of $10^{-12}$ \ergcms. All the flux ratios are calculated in the mentioned \chandra\ or \xmm\ band using the source peak count rate in the SBS (after conversion to the appropriate band). The complete catalog is available in the electronic version of this publication.}
\label{tab:srccat_counterpart}
\begin{tabular}{ccccccccc}
\hline
\hline
SBS X   &   ACIS flux	    & HRC flux	    & EPIC flux	     & $F_{\textrm{SBS}}/F_{\textrm{ACIS}}$ & $F_{\textrm{SBS}}/F_{\textrm{HRC}}$& $F_{\textrm{SBS}}/F_{\textrm{EPIC}}$      &	Vmag & $\log F_X/F_V$ \\
        & (0.5--7 keV)      & (0.1--10 keV) & (0.2--12 keV)  &                          &           &                   &           &      \\
\hline
1	    & $0.003\pm0.002$   & --	        & $85.1\pm0.3$	 & $1\pm0.6\times10^{5}$    & --	    & $7.5\pm1.5$       & --	    &	-- \\
2	    & $50.3\pm0.1$	    & --	        & $130.0\pm0.1$	 & $2.3\pm0.3$	            & --	    & $2.8\pm0.3$       & --	    &	-- \\
3	    & --	            & --	        & --	         & --	                    & --	    & --                & 10.77	    &	-2.25 \\
4	    & --	            & --	        & --	         & --	                    & --	    & --                & --	    &	-- \\
5	    & $41.8\pm0.2$	    & --	        & --	         & $0.6\pm0.1$	            & --	    & --                & --	    &	-- \\
6	    & --	            & --	        & --	         & --	                    & --	    & --                & --	    &	-- \\
7	    & $1.52\pm0.03$	    & --	        & --	         & $2.0\pm1.0$	            & --	    & --                & 9.0	    &	-2.53 \\
8	    & --	            & $11.3\pm0.1$  & $264.0\pm0.5$	 & --	                    & $41\pm7$	& $2.6\pm0.4$       & --	    &	-- \\
9	    & $33.9\pm0.2$	    & --	        & $0.88\pm0.02$	 & $0.8\pm0.3$         	    & --	    & $66\pm21$         & --	    &	-- \\
10	    & $39.5\pm0.2$	    & --	        & $0.06\pm0.004$ & $9.7\pm1.9$	            & --	    & $13100\pm2760$    & --	    &	-- \\
\hline
\end{tabular}
\end{table*}

\begin{table*}
\centering
\caption{SBS Catalog Part IV: catalogue columns containing information on previous detections/studies, and the possible nature of each source. The complete catalog is available in the electronic version of this publication.}
\label{tab:srccat_nature}
\begin{tabular}{ccccl}
    \hline
    \hline
SBS X	& \swift\ 	                & Alt. ID	            & Type	    & Comment \\
\hline
1	    & J174702.6$-$285259	& SAX J1747.0$-$2853	& LMXB	    &  Burster, Transient NS-LMXB \citep{Werner04}  \\
2	    & J174621.1$-$284342	& 1E 1743.1$-$2843	    & LMXB	    &  Persistent NS-LMXB candidate \citep{Lotti16}     \\
3	    & J174627.1$-$271122	& HD 316264	            & Ecl-Bin	&  Algol type, Gaia Parallax 6.5838+/-0.1913 mas (distance of 150 pc) \\
4	    & J174430.5$-$274600	& IGR J17445$-$2747	    & LMXB	    &  Transient NS-LMXB, this work, \citet{Shaw20}   \\
5	    & J174042.8$-$281807	& SLX 1737$-$282	    & LMXB	    &  Persistent, Burster, NS-LMXB, UCXB candidate \citep{intZand02,intZand07} \\
6	    & J174240.0$-$274455	& GRS 1739$-$278	    & LMXB?	    &  Transient, BH-XRB candidate \citep[e.g.][]{TetarenkoB16}, companion B-star or K giant?    \\
7	    & J174230.3$-$284454	& HD 160682     	    & Ecl-Bin	&  Beta Lyra, Gaia Parallax 11.38+/-0.05 mas (distance of 88 pc), negligible optical loading   \\
8	    & J174354.8$-$294441	& 1E 1740.7$-$2942	    & LMXB?	    &  Transient, BH-XRB, LMXB vs HMXB unclear \citep[e.g.][]{TetarenkoB16}     \\
9	    & J174445.2$-$295045	& XMMU J174445.5$-$295044& LMXB	    &  Transient, Symbiotic XRB at ~3.1 kpc \citep{Bahramian14b}    \\
10	    & J174451.5$-$292042	& KS 1741$-$293   	    & LMXB	    &  Burster, Transient NS-LMXB \citep{DeCesare07}   \\
\hline
\end{tabular}
\end{table*}

\subsection{Detections due to optical loading}\label{sec:ol}
Some SBS detections are contaminated by optical loading by bright optical counterparts on the \xrt\ detector\footnote{\url{https://www.swift.ac.uk/analysis/xrt/optical_loading.php}}. We used the \xrt\ Optical Loading Calculator\footnote{\url{https://www.swift.ac.uk/analysis/xrt/optical_tool.php}} to estimate possible contamination of the estimated count rates (and subsequently flux), and thus determine whether we can be confident that real X-ray emission was detected from sources near bright stars. Detections that are likely due to optical loading are denoted in the catalog by ``OL'' under the ``Type'' column.  We also comment on the severity of optical loading in each case in the ``Comment'' column (Table~\ref{tab:srccat_nature}). We found that in 12 sources (SBS X16, X44, X45, X82, X85, X86, X130, X135, X184, X198, X775, X784), all or most detected X-ray events are likely caused by optical loading and in 1 source (SBS X71), while optical loading is possible, real X-ray emission from the source is not ruled out. We verify that in each of these sources, the X-ray hardness ratio is $<0.1$, indicating an extremely soft spectrum consistent with optical loading \footnote{see https://www.swift.ac.uk/analysis/xrt/optical\_loading.php}.  It is possible that these sources may also provide some soft X-ray photons typical of chromospheric activity, but they certainly do not provide hard ($\gsim1.5$ keV) X-ray photons.  We do not consider these sources further in this paper, bringing our solid X-ray detections down to 91 sources. 

\section{Results}\label{sec:results}
Using the methods described above, we detected new faint (and bright) X-ray transients in the Galactic Bulge, and studied variability and faint transient activity from other known sources. 
\subsection{Previously Classified Sources}
\subsubsection{X-rays from stellar chromospheric activity}
Sixteen of the 17 optically bright ($V\lesssim13$) counterparts showing genuine X-ray emission show soft  X-ray hardness ratios ($< 0.5$; see section \ref{sec:analysis} for definitions), consistent with nearby (low $N_H$) foreground stars where the X-rays are produced by chromospheric activity. We use $\log F_X/F_V$= 0 as a conservative maximum for chromospheric activity, following the empirical dividing line between ABs and CVs in globular clusters \citep{Verbunt08}. We find that these active stars vary from $\log F_X/F_V\sim0$ to $-3$ (Fig.~\ref{fig:sbscatalog}, right). All but SBS X103 have soft \xrt\  spectra. We quote some key information below, from the Two Micron All Sky Survey \citep[2MASS, ][]{Skrutskie06}, Gaia \citep{Gaia2018,Bailer-Jones18}, and Tycho \citep{Hog00} catalogues.

SBS X3 matches HD 316264, a $V$=10.77 Algol-type F8 eclipsing binary at 150 pc.

SBS X7 associates with an eclipsing binary Beta Lyrae system, HD 160682 ($V\sim$9), which is a G5V star at 88 pc.

SBS X20, or Swift J173931.3$-$290953, matches HD 316072, a G9III star at $V$=9.9, and is transient (factor 15$\pm$5 increase over \chandra), though it was detected in 3 SBS epochs.

SBS X31 matches HD 316308, a K0 star at $V$=9.6, detected by \chandra\ as CXO J174854.0$-$285930. It was only seen in one epoch, where it was 100 times brighter than the \chandra\ detection.

SBS X34 (Swift J174023.9$-$285647) matches AX J1740.4$-$2856, which was identified with the nearby (56 pc), spectrally normal M dwarf 2MASS J17402384$-$2856527 by \citet{Lutovinov15}. Optical flares up to $V=9$ (from quiescent $V=13$) were seen from 2MASS J17402384$-$2856527 in \integral\ Optical Monitor data. The $\log F_X/F_V$ ratio is at the top of the range for chromospheric activity.

SBS X46, aka 1SXPS J174215.0$-$291453 \citep{Evans14}, is shown by \citet{Shaw20} to be an M dwarf binary. Details of its X-ray emission are in \S~\ref{sec:srcsxps}.

SBS X47 matches CXOGBS J173826.1$-$290149 (aka CX7; \citealt{Jonker11}), which was shown to be a K0 pre-main-sequence star by \citet{Torres06,Hynes12}. 

 SBS X48 matches CXOGBS J173629.0$-$291028 (CX10), which is HD 315992, a G7 variable star without evidence of binarity \citep{Hynes12}. The SBS detection is 20$\pm$10 times brighter than \chandra, but it is detected in 3 SBS epochs.
 
1SXPS J174034.5$-$293744 (SBS X55) is Tycho 6839-501-1, $V$=11.26. \citet{Gaia2018} gives $T_{\textrm{ eff}}$=4974 K, $d$=180 pc. 

SBS X63 matches V* BN Sgr, an Algol-type F3V eclipsing binary at 340 pc. The \xrt\ data show  strong flaring compared to archival \chandra\ and \xmm\ quiescent measurements (factor $>$8), but it is detected in 5 (of 14) SBS epochs.

Swift J175041.4$-$291644 (SBS X79) is very soft (hardness ratio of 0), and matches
2RXP J175041.2$-$291644, CXOGBS J175041.1$-$291644 (CX183, \citealt{Jonker11}, a factor $>$13 brighter), and 
HD 162120, an A2V star with $V$=8, at $d$=177 pc \citep{Hynes12,Gaia2016}.
  It is unusual to see X-rays from an A star; typically X-rays from A stars are thought to be produced by companion stars \citep{Schroder07,Hynes12}, which is consistent with the especially low $\log F_X/F_V$ and soft spectrum here.

Swift J174142.0$-$283321 (SBS X92) matches 2RXP J174141.9$-$283324 and 2CXO J174141.7$-$283324, associated with a $V$=13 star at 111 pc, with $T_{\textrm{eff}}=4139$ K, $L=0.13 L_{\odot}$ and $R=0.69 R_{\odot}$ \citep{Gaia2018}, thus a nearby dwarf K star.  SBS detected it in one epoch, $>$5 times brighter than the \chandra\ detection.

 SBS X132 matches CXO J174446.0$-$274732, which matches a $V$=12.3 star. \citet{Gaia2018} indicates $d$=920$\pm$50 pc, $R$=5 \Rsun, $T_{\textrm{eff}}$=4900 K. 

SBS X190 matches CXOGBS J175020.7$-$302652  (CXB11), which matches a $G$=12.2 star at 540 pc, with 4500 K, $R$=3.2 \Rsun, L=3.8 \Lsun \citep{Gaia2018}. An optical spectrum shows that H$\alpha$ is filled in, as often seen in chromospherically active stars \citep{Skiff14}.

SBS X558 matches 2RXP J174046.7$-$283849 and 
CXOGBS J174046.5$-$283850 (CX469).
\citet{Wevers16} identify the \chandra\  source with a saturated star at \radec{2000}{17}{40}{46}{59}{-28}{38}{49}{80} with high likelihood. This matches a Gaia star with Rp=12.7, $T_{\textrm{eff}}$=3850 K, at $d$=95 pc. The peak SBS flux is $>$16 times brighter than the \chandra\ and \xmm\ fluxes, and it is only detected in one SBS epoch.

SBS X803 matches CXOGBS J175432.1$-$292824 (CXB8), which matches 
 Tycho 6853-1571, with $V$=11.2. \citet{Gaia2018} gives $R_P$=10.5, $T_{\textrm{eff}}$=4900 K, $R$=3.4 \Rsun, and $d$=466$\pm$15 pc. \citet{Wevers17} find that this star shows an optical spectrum consistent with a normal star. The peak SBS flux is 17$\pm$9 times the \chandra\ flux, though it is detected in 5 of 21 epochs.

Finally, Swift J174701.3$-$291308 (SBS X103) matches 2XMM J174700.7$-$291309 and the Tycho star TYC 6840-337, $V$=11.6, with a Gaia  distance of 1.0$\pm0.1$ kpc \citep{Bailer-Jones18}, $T_{\textrm{eff}}=4298$ K, $R=11.6 R_{\odot}$ and $L=41 L_{\odot}$ \citep{Gaia2018}, thus a K giant. Its hardness (0.67) is consistent with a 1 kpc distance and a hard intrinsic spectrum, as seen in RS CVn binaries at this high $L_X$ of $5\times10^{31}$ \ergs\ (e.g. \citealt{Heinke05a}).  Its log$(F_X/F_V)$ of -2.5 is consistent with chromospheric activity.

\subsubsection{Known X-ray binaries}\label{sec:xrbs}
SAX J1747.0$-$2853 (SBS X1) is a transient X-ray binary discovered by \bepposax\ in 1998 \citep{intZand98a}. Follow-up observations confirmed its transient nature, and detection of X-ray bursts showed it harbors a neutron star, and constrained the source distance to $7.5\pm1.3$ kpc \citep{Werner04}. In the SBS observations we observed a long outburst detected from the first epoch of the survey on 2017-04-13, until 2017-10-05 (epoch 14, last detection), with the X-ray luminosity reaching $2\times10^{37}$ \ergs. We observed an initial peak, followed by a sharp decline around MJD 57860 and variability in a relatively faint state for the next 200 days, and eventually decay to quiescence by 2017-10-19, with the X-ray luminosity fading below $<1\times10^{36}$ \ergs\ (Fig.~\ref{fig:lc_sample}).

1E 1743.1$-$2843 (SBS X2) is a persistent source, without clear evidence of its nature \citep[e.g.,][]{delSanto06}. A recent deep X-ray study of this system by \citet{Lotti16} has indicated that this system is likely a neutron star LMXB. 1E 1743.1$-$2843 was detected persistently throughout our survey with a variable X-ray flux between $8\times10^{-11}$ \ergcms\ and $2.2\times10^{-10}$ \ergcms\ ($L_X$ of $6\times10^{35}$--$1.6\times10^{36}$ \ergs\ if at 8 kpc). 

\srcigr\ (SBS X4) is a transient discovered by \integral\ \citep{Bird07,Bird10}. \citet{Landi07} identified a \xrt\ counterpart with a 2--10 keV flux of $1.5\times10^{-13}$ \ergcms. \xrt\ has detected it in outburst twice since discovery. In April 2017, \integral\ detected a type I X-ray burst from it, indicating the compact object is a neutron star \citep{Mereminskiy17b, Mereminskiy17a}. Our SBS \xrt\ observations allowed a more precise localization to $\rmn{RA}(2000)=17^{\rmn{h}}~44^{\rmn{m}}~30\fs3$, $\rmn{Dec.}~(2000)=-27\degr~46\arcmin~00$, with uncertainty 2.2 arcseconds \citep{Kennea17}. \citet{Chakrabarty17}  obtained the most accurate \chandra\ localization of $\rmn{RA}(2000)=17^{\rmn{h}}~44^{\rmn{m}}~30\fs4$, $\rmn{Dec.}~(2000)=-27\degr~46\arcmin~00$, with uncertainty of 1 arcsecond. This allowed identification and study of the infrared counterpart \citep{Fortin17,Shaw17}. \citet{Shaw20} show from {\it Gemini} near-IR spectroscopy that the companion is a giant, estimating a distance of 2--8 kpc, and argue that \srcigr\ is a symbiotic X-ray binary where the neutron star has a low $B$ field, allowing X-ray bursts.

\begin{figure}
\includegraphics[scale=0.6]{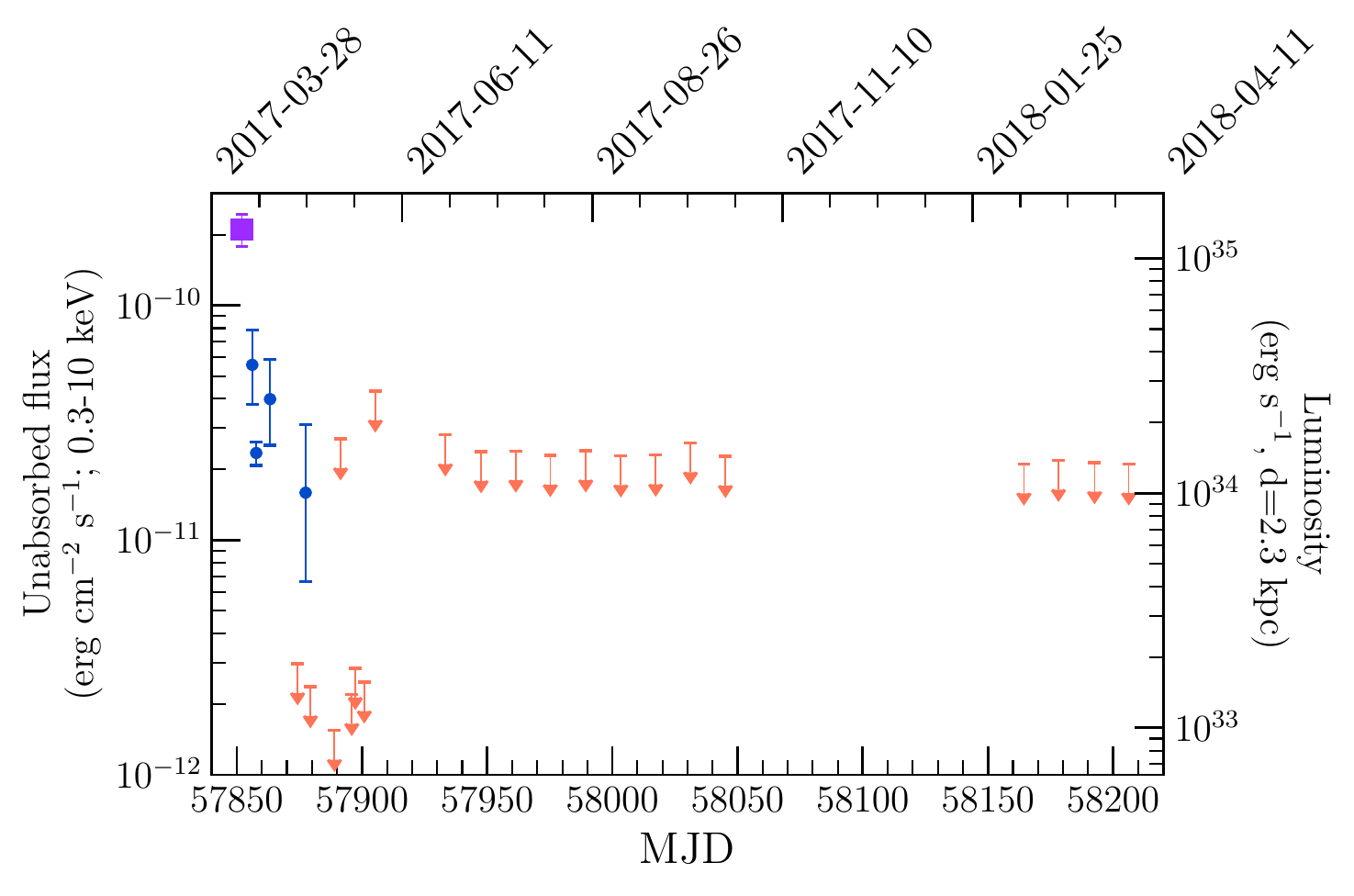}
\caption{X-ray light curve of \srcigr\ during year 1 of SBS. Blue circles and orange upper limits are based on \xrt\ observations from  this survey, plus target of opportunity observations. The purple square is the average source flux, as measured by \integral\ between April 7th and 11th \citep{Mereminskiy17b}. The source showed a decline over a period of 6 weeks. 
} \label{fig:igr1744}
\end{figure}

\srcigr\ was detected only in epochs 1-3 of SBS in 2017-18,  showing an overall decline to quiescence over a period of $\sim6$ weeks after the \integral\ observations, and strong variability on timescales of 
a few days (Fig.~\ref{fig:igr1744}). \xrt\ count rates were converted to flux assuming the best-fit model (absorbed power-law) as reported by \citet{Mereminskiy17b}. The detection on 2017-05-04 was preceded and succeeded by non-detections on 2017-05-01 and 2017-05-06. The highest X-ray flux measured in 2017 was from April 7th/11th \integral\ observations  \citep{Mereminskiy17b}. For a distance of 2.3 kpc \citep{Shaw20}, this  suggests a VFXT outburst with a peak $L_X$ of $\sim2\times10^{35}$ \ergs. However, the distance to the source is highly uncertain \citep{Shaw20}, and the full range of possible luminosities for a distance in the 2--8 kpc range is $L_X=1.5\times10^{35}$--$2.4\times10^{36}$ \ergs.

SLX 1737$-$282 (SBS X5) is a persistent, relatively faint X-ray source. \citealt{intZand02} discovered X-ray bursts from this system, confirming its neutron star nature. The faint X-ray luminosity of the system suggests that it may be an ultracompact X-ray binary \citep{intZand07}. SLX 1737$-$282 shows suggestive evidence of variability in the SBS, with X-ray luminosity between $9^{+8}_{-5}\times10^{34}$ \ergs\ and $5.6\pm1.2\times10^{35}$ \ergs, assuming a distance of 7.3 kpc \citep{Falanga08}. While the upper bound is consistent with previous studies of this source, our lower bound establishes the faintest level at which it has been observed \citep[Fig.~\ref{fig:lc_sample}; cf.][]{ArmasPadilla18}.

GRS 1739$-$278 (SBS X6) is a known BH XRB (aka V* V2606 Oph), that has undergone outbursts in 1996 \citep{Borozdin98} and 2014, the latter followed by several mini-outbursts \citep{Mereminskiy17_GRS}. It erupted again in 2016, unusually staying bright into 2018 \citep{Parikh18}. Our lightcurve of GRS 1739$-$278 shows variability by a factor of $\gsim5$ (Fig.~\ref{fig:lc_sample}).

1E 1740.7$-$2942 (SBS X8) is a well-known persistent BH (candidate) X-ray binary  \citep[][and references therein]{Sunyaev91, Natalucci14}. During our survey, it showed a variable X-ray luminosity between 2 and $7\times10^{36}$ \ergs, assuming a distance of $\sim$8.5 kpc \citep{Natalucci14, TetarenkoA20}. This is consistent with the system luminosity observed in the hard state \citep[e.g.,][]{Nakashima10}.

KS 1741$-$293 (SBS X10) is a known bursting neutron star LMXB \citep{intzand91, DeCesare07}. This system sometimes shows short subluminous outbursts with L$_X < 10^{36}$ \ergs\ \citep[e.g.,][]{Degenaar12b}. We detected a bright (L$_X > 10^{36}$ \ergs) outburst from KS 1741$-$293, with an X-ray luminosity averaging $\sim2\times10^{36}$ \ergs, and reaching $\sim10^{37}$ \ergs (assuming a distance of 8 kpc, a hydrogen column density of N$_H$ of $1.7\times10^{23}$ cm$^{-2}$, and a power-law photon index of 1.6, \citealt{Degenaar08}) in the third epoch. The source was not detected in the fourth epoch or later, with a (conservative) upper limit of L$_X < 1.5\times10^{36}$ \ergs\ (Fig.~\ref{fig:lc_sample}).

1A 1742$-$294 (SBS X11) is a bright persistent burster LMXB \citep{Pavlinsky94,Lutovinov01}. We detect this source in all epochs, with $L_X$ between $\sim1$ and $5\times10^{37}$ \ergs, assuming a distance of 8 kpc, N$_{\mathrm{H}}$ of $\sim1\times10^{23}$ cm$^{-2}$, and a power-law photon index of 1.6 \citep[e.g.,][]{Wijnands06}.

SLX 1744$-$299 (SBS X12) and SLX 1744$-$300 (SBS X14) are two bright persistent LMXBs. Detection of bursts from both systems indicated they are NS-LMXBs \citep{Skinner90, Pavlinsky94}. Given the small separation between these two sources ($\approx 2.6'$), SBS provides the first long-term monitoring of these sources in which they are distinguished from one another. We find both systems variable in the SBS (Fig.~\ref{fig:lc_sample}), with L$_X$ between $1.7\times10^{36}$ and $1.2\times10^{37}$ \ergs\ for SLX 1744$-$299, and between $2.7\times10^{36}$ and $1.4\times10^{37}$ \ergs\ for SLX 1744$-$300, assuming a distance of 8 kpc, hydrogen column density of $4.5\times10^{22}$ cm$^{-2}$, and a power-law photon index of 2.0 for both systems \citep[from \xmm\ spectral fitting of these sources,][]{Mori05}.

AX J1745.6$-$2901 (SBS X30) is a transient, eclipsing, bursting LMXB near the Galactic Centre, included in \xrt\ daily monitoring \citep[e.g.][]{Degenaar12,Ponti15,Ponti18}. We detected it in the May 18 and June 1, 2017 epochs, peaking at $L_X\sim5\times10^{34}$ \ergs, consistent with \citet{Degenaar17atel}.

GRS 1747$-$312 (SBS X60) is a transient, eclipsing, bursting LMXB in the globular cluster Terzan 6 \citep{intZand99,intZand03}. This source typically shows outbursts every $\sim6$ months. It is possible that there are multiple transient LMXBs residing in Terzan 6 which are not resolved by \xrt\ (given its high encounter rate, \citealt{Bahramian13}).  An outburst detected by {\it Suzaku} did not show an eclipse at the predicted time, which might indicate a second LMXB in the cluster \citep{Saji16}, though \citet{Vats18} argue that it could be the same LMXB. We confirmed the outburst detected in our survey was from GRS 1747$-$312 by performing a follow-up \xrt\ observation  at the time of one predicted eclipse ingress  \citep{intZand03}, which successfully detected the eclipse.

GRO J1744$-$28 (SBS X334) is a transient, bursting, pulsing LMXB \citep{Finger96} and is known to display low-level activity between its bright outbursts and quiescence \citep{Degenaar12b}. We detect it on June 29, 2017 at 20 times above its quiescent flux, as measured by \xmm\ and \chandra\ \citep{Wijnands02e}. This may be a continuation of the February outburst reported by \citet{Mereminsky17c,Sanna17,Russell17}.

\subsubsection{Known cataclysmic variables (CVs)}

SBS X76, or Swift J174015.9$-$290331,
matches AX J1740.2$-$2903, an {\it ASCA} source  \citep{Sakano00}. 
Several authors measured X-ray and optical   periods between 623 and 628 s \citep{Farrell10,Halpern10_atel}.
\citet{Thorstensen13} measured an orbital period (via radial velocities) of 0.2384 days, securely classifying this system as a DQ Her, or IP, CV. \citet{Farrell10} measure $N_H=4\times10^{21}$ cm$^{-2}$ from the \xmm\ observation, implying a distance below 4 kpc (using the extinction/distance relation of \citealt{Schultheis14}).

SBS X29 matches CXOGBS J174133.7$-$284033 (CX21, \citealt{Jonker11}), which was classified as a quiescent CV below the period gap by \citet{Britt14} and \citet{Wevers17} on the basis of strong optical flickering and an optical spectrum, and argued to be closer than 500 pc.

 SBS X35 matches CXOGBS J174009.1$-$284725 (CX5), and AX 1740.1$-$2847, which shows an X-ray periodicity of 729 s \citep{Sakano00}. \citet{Kaur10} identified a low-mass stellar counterpart, which \citet{Britt13} verified spectroscopically as an IP with a likely 125-minute orbital period.

SBS X146 matches CXOGBS J174028.3$-$271136 (CX128). \citet{Torres14} discovered strong H$\alpha$ and H$\beta$ emission lines in the optical spectrum of this source, along with several helium lines. They also detected broad emission lines from the Paschen series with complex profiles, and suggested the system is likely a CV. Assuming a power-law photon index of 1.8, N$_H$ of 10$^{21}$ cm$^{-2}$, and a distance of 1.0 kpc \citep{Torres14}, we estimate an X-ray luminosity of $1.7\times10^{32}$ \ergs, more than 10 times higher than the reported luminosity by \citet{Torres14}. It is detected in only one of 5 SBS exposures covering its location.

SBS X133 matches XMMU J175035.2$-$293557, which was identified as a likely IP by \citet{Hofmann18}
due to its X-ray spectrum and 511 second X-ray periodicity, interpreted as the spin period of the system. Our measured \swift\ flux is at least 50 times brighter than an upper limit we estimate from the quiescent \chandra\ flux, but is very close to the \xmm\ flux of \citet{Hofmann18}. This suggests that SBS X133 is one of the rare IPs that occasionally go into off states \citep{Kennedy17}.

\subsubsection{Likely symbiotic X-ray binaries or symbiotic stars}
Swift J175233.3$-$293944 (SBS X707) is consistent with the position of the \chandra\ Galactic Bulge Survey source CXB12 (or CXOGBS J175233.2$-$293944, \citealt{Jonker14}). Our \swift\ detection, at $F_X$=$2.5\times10^{-12}$ \ergcms, is $\sim$16 times brighter than the \chandra\ and \xmm\ detections, but with large errors.
\citet{Wevers16,Wevers17} identified CXB12 with a  $r'$=12.35 star (blended with a brighter neighbour $\sim$0\farcs5 away), which is also detected in \xmm\ Optical Monitor UV images (e.g. uvw1=14.57$\pm$0.01). \citet{Wevers17} used VLT/FORS2 spectroscopy to find an H$\alpha$ absorption line of EW=1.03($\pm0.05$) \AA. 
The ASAS V-band lightcurve \citep{Pojmanski97} shows variability of 0.2 magnitudes. \citet{Wevers17} use the fitted $N_H$ of an \xmm\ observation ($4.9\times10^{21}$ cm$^{-2}$) to infer a distance of 1.5-4 kpc, using the reddening map of \citet{Green15}, and thus a likely giant companion. The UV detection implies $L_{UV}=2.5\times10^{34}$($d$/2 kpc)$^2$ \ergs, while the \chandra\ and \xmm\ flux values imply  $L_X\sim1-2\times10^{32}$($d$/2 kpc)$^2$ \ergs. \citet{Wevers17} suggest this is likely a symbiotic system, with either a WD or NS primary. Our \swift\ measurement of $L_X\sim3\times10^{33}$ ($d$/2 kpc)$^2$ \ergs\ strongly favors the symbiotic explanation over the alternative RS CVn interpretation considered by \citet{Wevers17}. We also favor a WD, rather than NS, nature for the accretor, based on the high UV luminosity, naturally produced in boundary layers on WDs accreting at high rates \citep[e.g.][]{Patterson85,Mukai17}.

\srcigr\ (SBS X4) is a candidate symbiotic X-ray binary, also discussed in \S~\ref{sec:xrbs} above. Optical/IR follow-up of this object is discussed in \citet{Shaw20}.

XMMU J174445.5$-$295044 (SBS X9) is a rapidly variable X-ray transient \citep{Heinke09c}. K-band spectroscopy of the near-infrared counterpart indicated that it is a symbiotic X-ray binary at a distance of $\sim3.1$ kpc \citep{Bahramian14b}. This system was briefly detected in the first few epochs of the SBS, with peak X-ray luminosity of $1.4\times10^{35}$ \ergs.

\subsubsection{Young stellar cluster}
Swift J174550.3$-$284920 (SBS X101) matches the Arches Cluster, a young cluster of high-mass stars near the Galactic Centre, including multiple Wolf-Rayet stars \citep{Figer02}, that is a known X-ray source \citep{YusefZadeh02,Law04}. Detailed analysis of our \swift\ data on this source, and of archival \swift\ and \chandra\ data, is performed in Kozynets et al. (in prep). 

\subsubsection{Active galactic nuclei (AGN)}
SBS X18 matches GRS 1734$-$292, a known AGN with bright X-ray emission. Multiple broad optical emission lines at a redshift of 0.0214 verify its AGN nature \citep{Marti98}.  GRS 1734$-$292 is a variable X-ray source, with an observed range of $5$ - $7 \times 10^{-11}$ \ergcms\ in the 2--10 keV band \citep{Guainazzi11,Tortosa17}. We persistently detect GRS 1734$-$292 in the SBS with flux between $3\times10^{-11}$ and $9\times10^{-11}$ \ergcms\ in the 2--10 keV band, consistent with previous observations.

\subsubsection{Radio pulsars and pulsar wind nebulae}
SBS X13 matches G359.23--0.82, also known as the Mouse pulsar wind nebula, which is a  well-studied extended X-ray and radio nebula \citep{YusefZadeh87, Gaensler04, Mori05, Klingler18} associated with the radio pulsar PSR J1747$-$2958. Our peak absorbed flux (7$\pm3\times10^{-12}$ \ergcms) and average flux (7$\pm2\times10^{-12}$ \ergcms) are consistent with the catalogued \xmm\ flux (9$\times10^{-12}$ \ergcms), as expected for such a source. 

SBS X112 matches CXOGCS J174722.9$-$280904,  a known X-ray source catalogued by both \chandra\ and \xmm. Based on its X-ray and infrared properties, \citet{lin12}
classify this source as a rotationally powered pulsar. We marginally detect this hard source (with a hardness of 1.0) in the final stacked image of our survey, with an absorbed flux of $2_{-1}^{27}\times10^{-12}$ \ergcms.

\subsection{New or newly classified sources}
\subsubsection{Candidate CVs or X-ray binaries}
SBS X54 matches CXOGBS J175553.2$-$281633 (CX137, \citealt{Jonker11}). The OGLE-IV survey identified a very bright optical counterpart to CX137 with $I=15.11$ and $V-I=1.32$, and a sinusoidal modulation with a period of 0.43 days, interpreted as the orbital period \citep{Udalski12}. Based on the optical properties, \citet{Torres14} estimated  a lower limit distance of $>0.7$ kpc, and a lower limit on $L_X\geq5.8 \times 10^{30}$ \ergs. 
\citet{Bailer-Jones18} gave a Gaia distance estimate of 880$\pm60$ pc for the associated star. \citet{Torres14} suggested that this system is either a CV accreting at a low rate, or a quiescent LMXB. Assuming a power-law photon index of $\Gamma=1.8$, and N$_H$ of 10$^{21}$ cm$^{-2}$, we estimate a peak X-ray luminosity (in a single SBS detection) of $1.9\times10^{32}$ \ergs\ at 880 pc. This is 20 times brighter than the luminosity reported by \citet{Torres14}. \citet{Gomez20} recently reported on modeling of the optical light curve of this source and classified the source as a CV.

SBS X97 matches XMMU J174654.1$-$291542, which was discovered and discussed by \citet{Degenaar12b}, who identified strong spectral variations (large changes in the 4--10 keV flux, without corresponding changes in the 1-4 keV flux). 
\citet{Degenaar12} also suggested it may be associated with a Spitzer infrared source, with a 3.6 micron magnitude of 10.7.  All \swift\ photons are hard, consistent with the high $N_H$ (2-3$\times10^{22}$ cm$^{-2}$) measured by \citet{Degenaar12} from \xmm\ and \chandra\ data. If the Spitzer association is correct, this points towards a symbiotic system and the strong spectral variability indicates that an AGN nature is less likely.

SBS X374 matches 3XMM J174417.2$-$29394 with a subgiant optical counterpart, which \citet{Shaw20} argue has a WD companion in a 8.71 day period, accreting from a focused wind. We present further X-ray follow-up, supporting this scenario, in \S~\ref{sec:srcxmm} below. 

SBS X657, a hard source (hardness ratio of 0.67), matches CXOGBS J175359.8$-$292907 (CXB2).  The secure optical counterpart has a broad, variable H$\alpha$ emission line, and shows 
eclipsing or dipping on a 0.447-day period \citep{Wu15}. \citet{Wu15} also found possible optical evidence of an X-ray burst (indicating it is a candidate NS-LMXB), and estimated the distance to be 1-4 kpc, suggesting the detected \chandra\ $L_X$ corresponds to $10^{32}-4\times10^{33}$ \ergs. This source was also detected by {\it ASCA} as AX J1754.0$-$2929 \citep{Sakano02}, but was not detected by \rosat, which could be due the hard and absorbed spectrum of the source (compared to the soft band of the \rosat\ imager).

\subsubsection{Sources with optical counterparts of unknown nature}
SBS X39 (intermediate hardness of 0.4) appears to be associated with the $P_{\textrm{orb}}$=0.282	day eclipsing binary  OGLE BLG-ECL-116765 \citep{Soszynski16}, with $G$=17.9. This star is $2.0^{+1.5}_{-0.5}$ kpc away \citep{Bailer-Jones18}, so should be somewhat reddened. The peak SBS X-ray flux (in 2 SBS epochs) at a 2 kpc distance gives an X-ray luminosity of $L_X\sim2\times10^{33}$ \ergs (150 times higher than the \chandra\ upper limit), and the source shows a $\log F_X/F_V$ value of 1.1,  suggesting that this may be an X-ray binary. It is also possible that the optical association is spurious.

SBS X78, CXOGBS J175316.4$-$283812 (CX46), was identified by \citet{Wevers16} at high confidence with a star that was saturated in their photometry ($r'<12$). This star has a Gaia-measured $T_{\textrm{eff}}=4058$ K, at a distance of $1.5\pm0.2$ kpc \citep{Bailer-Jones18}, and an inferred radius of 31 \Rsun; thus, a K giant. 
The X-ray hardness ratio of 0.5 is consistent with the foreground nature of this source. The $\log F_X/F_V$ suggests that this object may be an RS CVn star, but the peak $L_X\sim3\times10^{32}$ \ergs\ is unusually large for that type of system, hinting it may be a symbiotic star. Follow-up spectroscopy of this reasonably bright star may be necessary to ascertain its nature.

SBS X109, a hard source (hardness ratio of 0.7), matches CXOGBS J175249.3$-$284009 (CX164).  \citet{Wevers16} found a high confidence counterpart, with $r'=19.2$, $i'=17.3$, and near-infrared magnitudes $K_S$=12.84, $H$=13.24, $J$=14.31 from the VISTA Variables in the Via Lactea survey \citep[VVV, ][]{Minniti17}. The reddening-free parameter $Q=(J-H)-1.70(H-K_S)$ gives $0.39\pm0.04$, which is close to that expected for late-type (K-M) stars \citep{Comeron05,Negueruela07}.  We can place limits on the reddening from the range of intrinsic colors of giant stars, compared to the observed $J-K$=1.47; we measure $0.27<E(J-K)<0.9$, $2.7\times10^{21}<\mathrm{N}_{\mathrm{H}}<9\times10^{21}$ cm$^{-2}$, $0.11<A_K<0.37$.
If this star is a K0 giant, we can estimate its distance as 6-7 kpc; a later spectral type would give a larger distance.  This distance range indicates $L_X>10^{34}$ \ergs, suggesting accretion, possibly in a symbiotic X-ray binary.

SBS X129, a hard source (hardness ratio of 0.9), matches CXOGBS J174151.2$-$270223 (CX11). \citet{Wevers16} find a high-likelihood optical counterpart, with $i'$=18.9, $r'$=21.1. \citet{Britt14} find this star to be variable. \citet{Greiss14} give $K$=13.5 for this star.

SBS X136, a moderately hard \swift\ X-ray source (hardness 0.63) matches CXOGBS J175007.0$-$300154, or CX124.
\citet{Udalski12} identify CX124 with OGLE BUL-SC5 134677, a $V=14.6,I=12.5$ star showing irregular variability.
\citet{Wevers16} confirm this identification with high likelihood. Gaia gives a 1.9$\pm0.2$ kpc distance for this star, indicating a peak $L_X$  $\sim6\times10^{32}$ \ergs\ (however with large errors). 

SBS X148, a fairly hard source (hardness 0.63) matches CXOGBS J173935.7$-$274023, or CX342. 
\citet{Wevers16} identify a high likelihood counterpart, with $i'$=20.6, $r'$=22.4. \citet{Greiss14} identify the NIR counterpart, with $K$=16.5.

SBS X180 matches CXOGBS J174718.2$-$304110 (CX34), and is a hard (hardness ratio of 0.9) source. \citet{Greiss14} find a 2MASS match with $K$=12.3, which has a low probability ($3\times10^{-3}$) of a false match.

SBS X210, a hard source (hardness ratio of 0.8), matches CXOGBS J175414.5$-$282150 (CX123). \citet{Wevers16} find a likely counterpart with $i'$=18.8, and colors consistent with a reddened, distant object. The \chandra\ position is also consistent with a bright (8.3 mag) 24-micron MIPSGAL source \citep{Gutermuth15}.

SBS X267, a hard source (hardness ratio of 0.9), matches CXOGBS J173440.8$-$291930 (CX242). \citet{Wevers16} found a likely counterpart with $i'$=19.0, $r'$=20.9.

SBS X417 has hardness 0.5, and matches 3XMM J174938.6$-$290333, as well as a Gaia source with $G$=13.2, $T_{\textrm{eff}}$=3900 K, $R$=9.2 \Rsun, at $d=1.58\pm0.13$ kpc. This matches the AAVSO variable source GDS J1749386$-$290333 \citep{Watson06}.

SBS X557, or Swift J174217.6$-$285651, is a hard source (1.0), which matches 2RXP J174217.9$-$285649 and 3XMM J174217.8$-$285647, and is seen in \chandra\ ObsID 18995, at flux levels within $<$factor of 10 of our \swift\ detection. The \xmm\ position matches 2MASS 17421780$-$2856478 ($K$=7.1, $V$=12.9), which has a Gaia distance \citep{Bailer-Jones18} of 3.2$^{+1.6}_{-0.8}$ kpc, and $T_{\textrm{eff}}$ of 3460 K, indicating a late-type giant. Its SBS flux implies $L_X=1.1\times10^{33}$ ($d$/3.2 kpc)$^2$ \ergs, which is exceptionally high for RS CVn systems, indicating a  probable symbiotic system. 

SBS X669 is a soft (0.1) source, matching CXOGBS J173908.3$-$282041 (CX126), the \rosat\ source 1RXH J173908.2$-$282040, and 2MASS 17390834$-$2820405 ($K$=10.2). \citet{Greiss14} find a high-likelihood $K$=11 star. \citet{Wevers16} find a high likelihood counterpart, saturated, with H$\alpha$=13.8. Our \swift\ detection (in 2 epochs) is a factor of at least 16 times brighter than the quiescent \chandra\ flux.

SBS X744, a hard (0.88) source, matches CXOGBS J173757.3$-$275213 (CX167), and is $>$8 times brighter in the single SBS detection.  
2MASS 17375732$-$2752137	($K$=8.4, $J$=9.7) is associated with it by \citet{Greiss14} with high probability. This matches a Gaia star with $R_p$=11.4, $T_{\textrm{eff}}$=4000 K, $R$=7.5 \Rsun, at 920$\pm$40 pc. It is also variable in the UV (Rivera~Sandoval et al., in prep.) This may be an RS CVn, or a symbiotic star.

SBS X771, a hard (0.67) source, matches CXOGBS J175737.9$-$280953 (CXB19).
\citet{Wevers16} identify a $r'$=17.8, $i' \sim$14.6 (saturated) very red source, with high likelihood. 
\citet{Greiss14} find VVV J175737.99$-$280952.94 to be a match at high likelihood, with  $K$=12.9, $J$=14.
A Gaia star is associated, with $R_p$=15.9, $d=4.5^{+3.7}_{-1.9}$ kpc. 
This is a candidate for a symbiotic star, given the peak X-ray luminosity of $L_X > 1.2\times10^{33}$ \ergs.

\subsubsection{Sources without clear optical/IR counterparts}
We attempt to classify most sources without optical/IR counterparts using their hardness ratio and variability.

The following objects, though they lack clear optical counterparts, have soft X-ray colors (quoted in parentheses, as the fraction of photons above 1.5 keV) indicating their foreground (and thus, likely chromospherically active) nature (see \S 3.1).

SBS X15 increased by a factor $>$200 from the \chandra\ upper limit, though it was only detected in the SBS stack. All detected photons are below 0.5 keV (so hardness is 0.0).

SBS X17 (0.25) increased by a factor $>12$ compared to the \chandra\ upper limit, and is only clearly detected in the SBS stack, but may have shown a flare. The detection lies 15" from the unidentified {\it ASCA} source AX J1735.1$-$2930, seen at $F_X=4.9\times10^{-12}$ erg cm$^{-2}$ s$^{-1}$ \citep{Sakano02}, which is similar to the peak SBS flux.

SBS X19 is also extremely soft (hardness 0), and highly variable (factor $>$62).
 
SBS X36 is a very soft (hardness ratio of 0.08) transient ($>$26 times brighter than \chandra\ upper limit).

SBS X37 is soft (0.18) and transient ($>$17 times brighter than \chandra\ upper limit).

SBS X38 is soft (0.13), without any \chandra\ or \xmm\ observation yet to measure its flux variation.
 
SBS X64 is a soft (0.3) transient (factor $>$110). 

SBS X113 is soft (0.18), with no \chandra\ observation of the region, but it matches the \rosat\  source 2RXP J174522.6$-$281732.

SBS X122 (CXOGBS J173508.2$-$292957, CX8) is soft (0.27), and not obviously varying. 2\arcsec\ away is the Tycho star TYC 6839-612, $V$=11.6, $d$=1.5 kpc, 4000 K, 22 \Rsun, 110 \Lsun. The \chandra\ detection is 4.8' off-axis, which might account for the positional offset of 2\arcsec\ from this counterpart.
 
SBS X546 is soft (0.4).

SBS X714 is soft (0.3), and matches the position of CXOGBS J175327.8$-$295718 (CXB51).
 
The following sources showed hard X-ray spectra, and variation by a factor of $\sim$5 or more, compared to previous \chandra\ and/or \xmm\ observations.
 
SBS X21 (Swift J175233.9$-$290952) is a mysterious hard (hardness ratio of 0.86) transient (factor $>$24 compared to \chandra\ upper limit) source. \citet{Shaw20} were unable to confirm an optical/IR counterpart, even though they used a DDT \chandra\ observation to obtain a precise localization during outburst. We discuss the X-ray data on this object in more detail in \S~\ref{sec:srcsw} below.

SBS X83 is a hard (hardness ratio of 0.86) transient (factor $>$6, compared to a \chandra\ upper limit).

SBS X91, a spectrally hard source (hardness ratio of 0.73), matches the hard \chandra\ source 2CXO J174333.9$-$283953 \citep{Wang16}, and requires variability by a factor $>$4.

SBS X131 matches CXOGCS J174622.0$-$285201 \citep{Muno06a}. It is hard (hardness ratio of 0.7), and varied by a factor of $>$30 between the \chandra\ and \swift\ fluxes.

SBS X140 is a hard (hardness ratio of 1.0) transient (varying by a factor $>$7 compared to the \chandra\ upper limits). It may be associated with the WISE YSO candidate J174310.86$-$270154.7	\citep{Marton16}.

SBS X195 is hard (hardness ratio of 0.86), and its SBS flux increased over the \chandra\ upper limit by a factor of at least 7.5.

SBS X738 (hardness ratio of 0.86) matches 2CXO J175700.5$-$280735 \citep{Evans19}.
The \swift\ detection is at least 36 times the detected \chandra\ flux.

The following sources showed fluxes within a factor of 5 of a previous \chandra\ or \xmm\ detection, and hard X-ray spectra. A relative lack of variability with a hard X-ray spectrum suggests either an AGN, or rarer hard stellar X-ray sources such as IPs, or persistent very faint X-ray binaries.  We checked the positions of these sources in the Very Large Array Sky Survey 2-4 GHz continuum images \citep{Lacy20}\footnote{https://science.nrao.edu/vlass/}, since AGN are often radio-bright, but did not find any radio counterparts to these sources.

SBS X22, a hard source (0.72), matches CXOGBS J175322.5$-$290535 (CX151), and is consistent with the \chandra\ flux. 
\citet{Wevers16} identified a potential optical counterpart (with $r'$=18.9), but the likelihood ratio for this object over alternative counterparts is $<$1, indicating the counterpart is likely to be spurious. 

SBS X24 matches CXOGBS J175253.0$-$292209 (CX17), is hard (0.65), and  consistent with the \chandra\ flux. 
 
SBS X26 matches CXOGBS J175244.5$-$285852 (CX201) and is fairly hard (0.56). \citet{Wevers16} identified a potential optical counterpart, but the likelihood ratio is $<$2, so it is likely spurious. 

SBS X27 matches with OGLE BLG195.7 120277, an RR Lyrae star, but RR Lyrae stars are not expected to be X-ray emitters. This is likely to be a spurious match. 

SBS X66, a hard source, has a \swift\ flux consistent with the flux previously seen by \xmm, \swift\ (the 1SXPS survey, \citealt{Evans14}), and \rosat. 

SBS X69 matches the \rosat\ source [SBM2001]  23 \citep{Sidoli01a}, and the \chandra\ source CXO J174216.9$-$283707 \citep{Evans19}. It is hard (0.75), and not obviously varying compared to the \chandra\ flux. 

SBS X114 matches CXOGBS J174656.0$-$303259 (CX35), and is hard (1.0). Its flux is  consistent with the \chandra\ flux.

SBS X124, a hard source (0.64), matches 3XMM J174917.7$-$283329. The \xmm\ flux may lie within a factor of 2 of the \swift\ flux. It may match the 2MASS and UKIDSS star UGPS J174917.66$-$283328.7, a $K$=11.9 source.

SBS X294, a hard source (0.83), matches CXOGBS J173808.8$-$292216 (CX136). There is no  optical counterpart with a strong likelihood of a match \citep{Wevers16}.
There is a possible association with the MIPSGAL 24 micron 8.4 mag point source MG358.7691+01.1559 \citep{Gutermuth15}, which likely matches GLIMPSE G358.7690+01.1557.

SBS X423 is hard (1.0), and matches 2RXP J175029.3$-$285954 
and CXOGBS J175029.1$-$290002 (CX13). It is persistent; the \swift\ flux is consistent with the \chandra\ and \xmm\ fluxes. There is no clear optical counterpart \citep{Wevers16}.

SBS X592 is hard (1.0), and consistent in position and flux with 3XMM J173918.2$-$292349.
It may correspond to MIPSGAL source MG358.8819+00.9293, which has a 24 micron magnitude of 7.1 \citep{Gutermuth15}.

SBS X731 is a hard (1.0) source. It is probably not associated with OGLE BLG-ECL-19927, which is 12\arcsec\ away, as that nearby active binary is unlikely to show a hard X-ray spectrum.

SBS X807,  a hard source (0.8), matches 2XMM J175516.6$-$293655 in position and flux.

\subsection{X-ray follow-up on three sources}
\subsubsection{\srcsw}\label{sec:srcsw}
\begin{figure}
\includegraphics[scale=0.6]{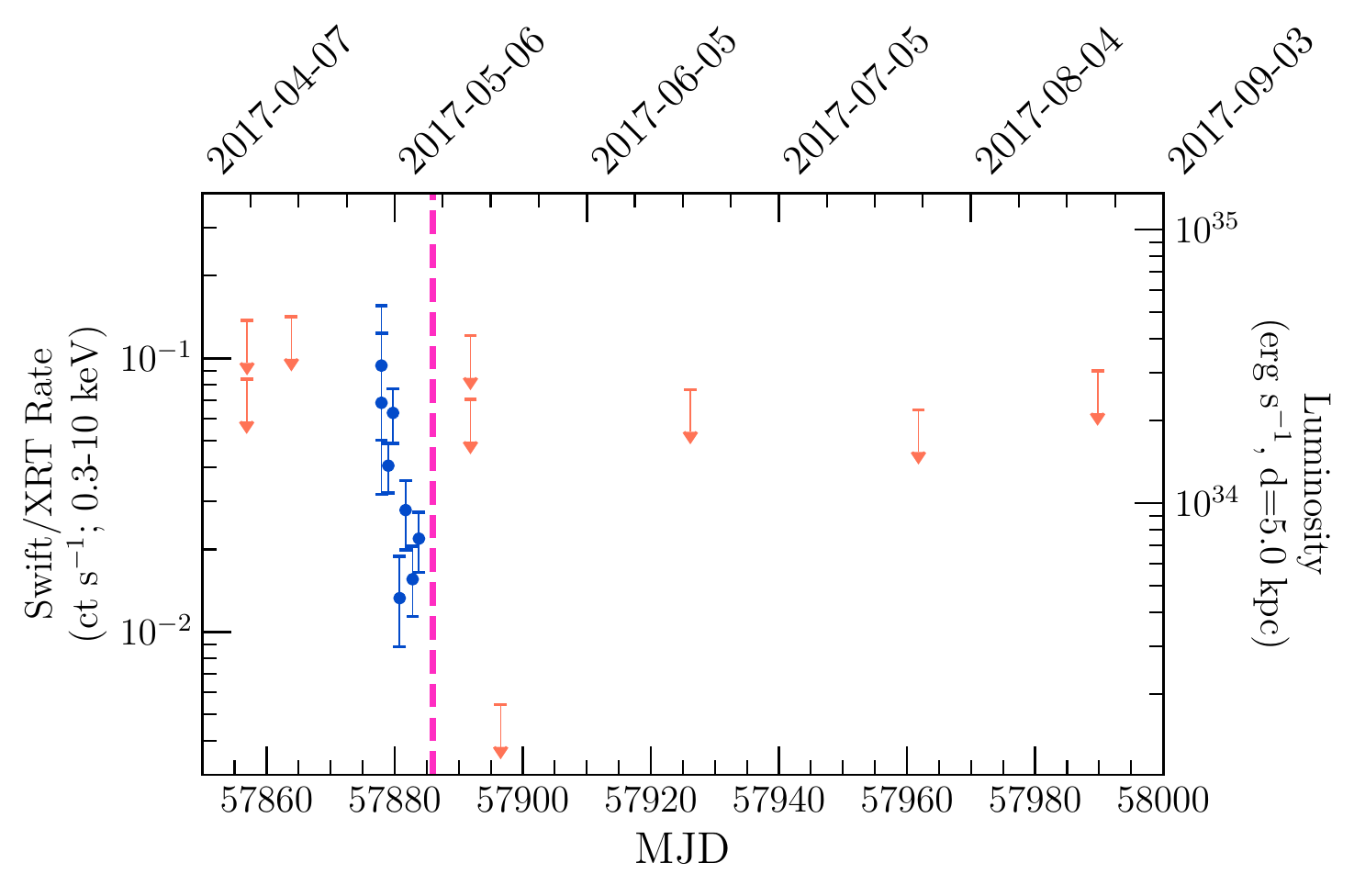}
\caption{\xrt\ light curve of \srcsw\ during the first year of SBS. \xrt\ count rates were converted to flux assuming the best-fit model (an absorbed power-law) fitting the \xrt\ spectrum (see text). Upper limits are  2-$\sigma$. The pink dashed line indicates the time of the VLA observation.} 
\label{fig:swj1752lc}
\end{figure}

\srcsw\ (SBS X21) was first discovered in epoch 3 of the SBS (on 2017-05-04) with a \xrt\ count rate of 0.05$^{+0.04}_{-0.02}$ ct s$^{-1}$ \citep[in the 0.5--10 keV band;][]{Bahramian17b}. We obtained a short \chandra\ observation on 2017-05-20 and identified a counterpart at
\radec{2000}{17}{52}{33}{903}{-29}{09}{47}{95}, with an uncertainty of $\sim$0.5 arcseconds \citep{Maccarone17a}.
Archival \chandra\ data  \citep{Jonker11} showed no prior detection, indicating a flux increase of a factor $>$24. Follow-up \xrt\ observations of \srcsw\  showed a steady decline over 2--3 weeks, with no later detections (Fig.~\ref{fig:swj1752lc}). 

\begin{figure}
\includegraphics[scale=0.3]{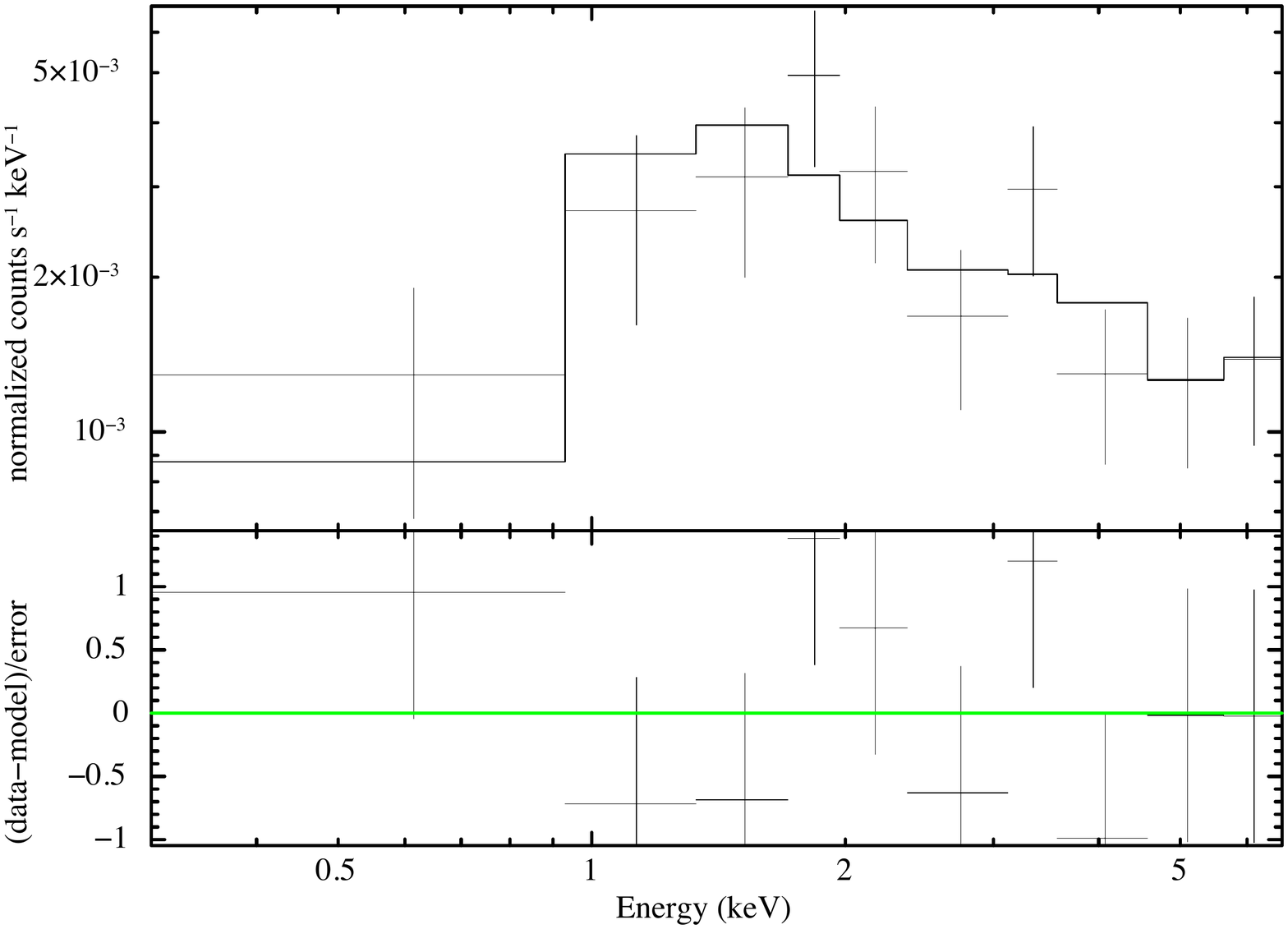}
\includegraphics[scale=0.3]{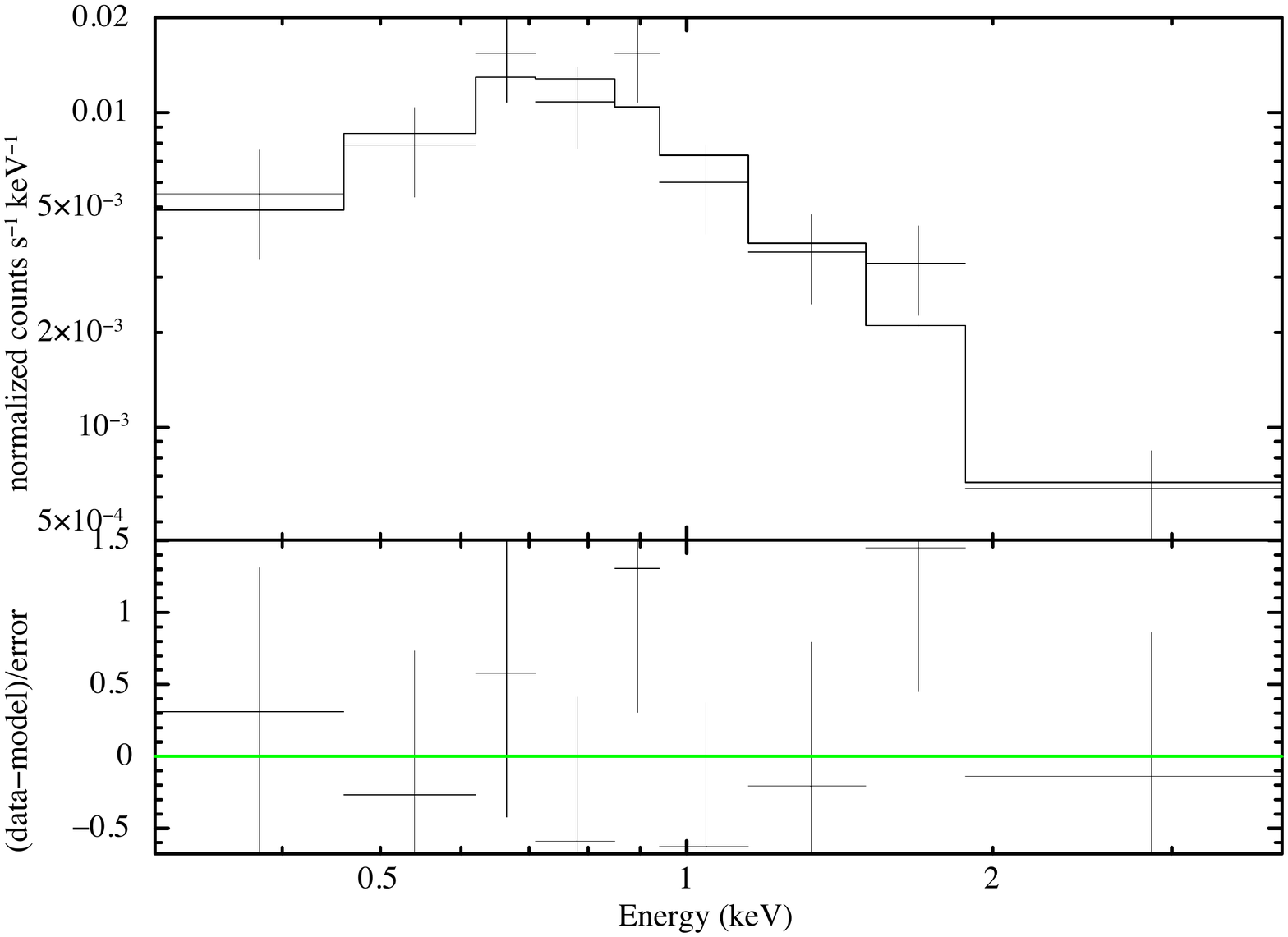}
\caption{\xrt\ spectra of \srcsw\ (top) and \srcsxps (bottom) during the first year of SBS and the best-fit models. Both spectra are rebinned ($\geq10$ counts per bin) for visualization here.} 
\label{fig:swift_specs}
\end{figure}

We extracted an X-ray spectrum from all \xrt\ observations in which the source was detected (all in PC mode) using the \xrt\ automated pipeline\footnote{\url{https://www.swift.ac.uk/user\_objects/}}, and performed spectral analysis in the 0.3--10 keV band using {\sc Xspec} \citep[version 12.10.1,][]{Arnaud96}, assuming \citet{Wilms00} abundances  and \citet{Verner96} photo-electric cross-sections (Fig.~\ref{fig:swift_specs}). The merged spectrum contains a total of $\sim100$ counts, thus we used W-statistics for minimization \citep{Cash79}.\footnote{See \url{https://heasarc.gsfc.nasa.gov/docs/xanadu/xspec/manual/node304.html}} Given the low quality of the spectrum, we modelled it with a simple absorbed power-law (TBABS$\times$PEGPWRLW in {\sc Xspec}). This model provides a W-statistic of 87.4 for 98 degrees of freedom, and a null hypothesis probability (the probability that the data is drawn from the fitted model, given the goodness and degrees of freedom) of 67\%. This fit indicates a hydrogen column density of $N_H \leq8\times10^{21}$ cm$^{-2}$, power-law photon index of $\Gamma=0.4\pm0.4$, and an average unabsorbed flux of $2.1_{-0.5}^{+0.6}\times10^{-12}$ \ergcms\ in the 0.5--10 keV band.\footnote{All uncertainties are at 90\% for a single interesting parameter.} The constraint on the hydrogen column density suggests that \srcsw\ is within $\lsim 5$ kpc, and the very hard power-law photon index indicates that the X-ray emission is at least in part caused by non-thermal processes (like inverse-Compton scattering in X-ray binaries with L$_X\sim$ 10$^{34}$ - 10$^{36}$ \ergs).

The \xrt\ spectrum hints at the presence of an emission line around $\sim$ 6.3-6.9 keV. Adding a gaussian emission line component to the power-law model (TBABS$\times$[PEGPWRLW+GAUSSIAN] in {\sc Xspec}) indicates a constrained line with a mean energy of $kT=6.6\pm0.3$ keV, a line normalization of $3.3_{-2.4}^{+3.4}\times10^{-5}$ (which is not highly significant, but inconsistent with zero), and the width of the gaussian corresponds to a standard deviation of $\sigma< 0.9$ keV. Adding this line also slightly changes the constraints on $N_H$ and photon index to $N_H=4.0_{-3.7}^{+8.3}\times10^{21}$ cm$^{-2}$ and $\Gamma=0.8_{-0.5}^{+0.6}$. However, the poor quality of the data does not allow us to robustly establish  whether the line is real.

A radio observation of \srcsw\ at 10 GHz by the Karl G. Jansky Very Large Array observatory on May 13th, 2017 (three days after the last X-ray observation, Fig.~\ref{fig:swj1752lc}) yielded a 3-$\sigma$ upper limit of $\sim$15 $\mu$Jy. This is more suggestive of a neutron star, rather than black hole, accretor if the source is an X-ray binary \citep{TetarenkoA17}.

\subsubsection{\srcsxps}\label{sec:srcsxps}
\srcsxps\  (SBS X46) was detected in multiple epochs of our 2017 survey \citep{Maccarone17b}. The combined \xrt\ data gives a position of  \radec{2000}{17}{42}{15}{14}{-29}{15}{01}{9}, with an uncertainty of 2.9$\arcsec$, matching \srcsxps\ from \citep{Evans14}, and the soft \rosat\ X-ray source 1RXS~J174216.4$-$291454 \citep{Sidoli01b}.
It is also clearly detected in 2016 \xmm\ observations.

We used \xmm\ spectra of the source from 2016-03-09 and 2016-08-30, produced by the \xmm\ Processing Pipeline System (PPS) and performed spectral analysis in the 0.2-12 keV band using \textsc{Xspec}. We fit spectra from all EPIC detectors simultaneously for each epoch, and included a cross-instrument calibration coefficient in the model. All \xmm\ spectra for this source contain $>200$ counts, so we used $\chi^2$ statistics. No single-component model (absorbed blackbody, disk-blackbody, power-law, comptonization, or diffuse emission from hot plasma) provided a satisfactory fit. We found a two-temperature hot gas model \citep[MEKAL in \textsc{Xspec},][]{Mewe86,Liedahl95} provides reasonable fits in both epochs (TBABS$\times$[MEKAL+MEKAL] in \textsc{Xspec}; Table~\ref{tab:xspsspec}).

We also extracted a spectrum from all the \xrt\ observations covering the source (performed between 2012 and 2018, all in photon-counting mode) and fit it with the same model as the \xmm\ data. The \xrt\ spectrum contains only $\sim100$ counts (Fig.~\ref{fig:swift_specs}). Thus we fit this data set using W-statistics for minimization. The two-temperature hot gas model fit the \xrt\ spectrum well, but it indicates a higher plasma temperature (Table~\ref{tab:xspsspec}). 

The \xrt\ flux is higher than that from \xmm, 
showing strong evidence for variability. However, evidence for variability within the \xrt\ data is weak due to the low signal-to-noise ratio in the individual \xrt\ exposures. \citet{Shaw20} identify a late M dwarf optical counterpart with a measured Gaia  \citep{Luri18} distance of $\sim$78 pc. This indicates that the source X-ray luminosity is variable between $1\times10^{29}$ and $2\times10^{30}$ \ergs, consistent with typical luminosities for nearby chromospherically active M dwarfs.

\begin{table*}
\centering
\caption{Best-fit parameter values for \xmm\ and \swift\ spectra of \srcsxps, fit with a two-temperature hot plasma (TBABS$\times$[MEKAL+MEKAL]). The \xmm\ spectra also include a cross-calibration coefficient between PN, MOS1 and MOS2. Norm$_\textrm{MEKAL}$ values use the \textsc{Xspec} model parametrization. F$_\textrm{X}$ is unabsorbed flux in the 0.3--10 keV band. $\chi^2_\nu$ represents reduced $\chi^2$, D.O.F represents degrees of freedom, and N.H.P. represents null hypothesis probability. All uncertainties are at 90\% confidence. The \xrt\ spectrum was fitted using w-stat statistics, thus $\chi^2_\nu$, D.O.F, and N.H.P are not reported for it.}
\label{tab:xspsspec}
\begin{tabular}{lccc}
\hline
\hline
Date                        &   2016-03-09                      & 2016-08-30                & 2012-2018                         \\
Telescope                   &   \xmm\                           & \xmm\                     & \xrt\                             \\
Obs. ID.                    &   0764190301                      & 0764191801                & --                                \\
\hline
N${_\text{H}}$ (cm$^{-2}$)  & $<9.4\times10^{19}$               & $<3.5\times10^{20}$       &  $<3\times10^{21}$                \\
kT$_1$ (keV)                & 0.3$\pm$0.01                      & 0.28$_{-0.04}^{+0.02}$    &  0.3$_{-0.08}^{+0.2}$             \\
Norm$_1$                    & 9.9$_{-0.5}^{+0.9}\times10^{-5}$  & 8$\pm2\times10^{-5}$      &  14$_{-7}^{+40}\times10^{-5}$     \\
kT$_2$ (keV)                & 1.01$\pm$0.05                     & 0.9$\pm0.1$               &  4$_{-2}^{+50}$                   \\
Norm$_2$                    & 7.7$\pm0.7\times10^{-5}$          & 6$\pm1\times10^{-5}$      &  19$_{-6}^{+7}\times10^{-5}$      \\
Calib. (MOS1/PN)            & 0.94$\pm$0.08                     & 1.1$\pm0.1$               &  --                               \\
Calib. (MOS2/PN)            & 0.97$\pm$0.08                     & 0.9$\pm0.1$               &  --                               \\
F$_\textrm{X}$ (\ergcms)    & 2.6$\pm0.1\times10^{-13}$         & 2.4$\pm0.2\times10^{-13}$ &  5.1$_{-1.1}^{+1.6}\times10^{-13}$\\
$\chi^2_\nu$/D.O.F          & 1.2/77                            & 0.99/49                   &  --                          \\
N.H.P. (\%)                  & 8.4                               & 48                        &  --                               \\
\hline
\end{tabular}
\end{table*}

\subsubsection{\srcxmm}\label{sec:srcxmm}
\srcxmm\ (SBS X374) was first detected in SBS on 2017-09-21, and is marginally detected in other epochs. This source, $\sim7$\arcmin\ from the bright X-ray binary 1E 1740.7$-$2942, is covered in multiple \swift, \chandra, and \xmm\ observations.  
\citet{Shaw20} noted the 8.69 day period from ASAS-SN monitoring, spectroscopically classified the visible star as a K subgiant, and used radial velocities to identify a dark companion with 1$>$M$>$0.4 \Msun, i.e. a WD. \citet{Shaw20} also used the mass ratio and observed X-ray luminosity to argue that the white dwarf accretes from the subgiant via a focused wind.  

We combined all relevant \swift\ observations to find an enhanced position of \radec{2000}{17}{44}{16}{92}{-29}{39}{43}{6}, with an uncertainty of 2\farcs3, consistent with previously catalogued sources by \xmm\ \citep[3XMM J174417.2$-$293944,][]{Rosen16}, \chandra\ \citep[CXO J174417.2$-$293943,][]{Evans10}, and \swift\ \citep[1SXPS J174417.2$-$293946,][]{Evans14}. We extracted spectra from all \chandra\ and \xmm\ observations in which the source is entirely located on an active detector chip in imaging mode. These include 7 \chandra/ACIS-I and ACIS-S observations and 2 \xmm/EPIC observations. The source is clearly detected in the \xmm\ observations in this sample with high signal-to-noise ratio, and contamination from 1E 1740.7$-$2942 appears to be minimal. Therefore, we acquired source and background spectra produced from the \xmm\ PPS for this source. In \chandra\ observations, the source is typically located at high off-axis angles. Thus, we used MARX \citep{Davis12} and ACIS-EXTRACT \citep{Broos10} to estimate the shape of the off-axis point-spread function and extract source and background spectra accordingly. Due to the low number of events per spectrum, we combined \chandra\ spectra from observations taken within days of each other using \texttt{combine\_spec} in \textsc{Ciao} \citep{Fruscione06}. We performed spectral analysis on every individual \chandra\ and \xmm\ spectrum separately in \textsc{XSPEC}, in the 0.3--10 keV and 0.2-12 keV ranges respectively. All spectra were binned to have at least 15 to 50 counts per bin, and we used $\chi^2$ statistics for spectral fitting and goodness-of-fit evaluation.

\begin{figure}
\includegraphics[scale=0.3]{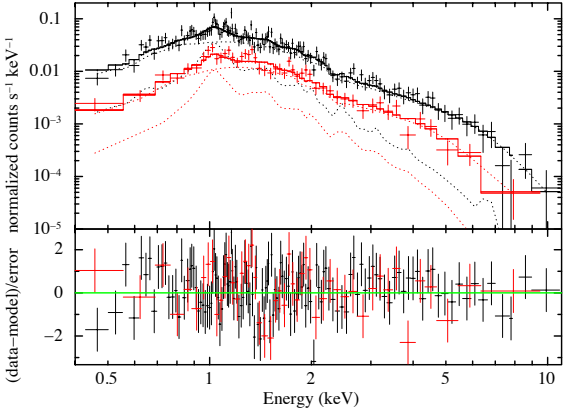}
\caption{Spectra of \srcxmm\ from 2012-09-07 (\xmm, PN in black, MOS2 in red) as fitted by MEKAL+power-law.}
\label{fig:xmmspec}
\end{figure}

As demonstrated in Table~\ref{tab:xmmspec}, a soft absorbed power-law fits most of the data reasonably well. However, there are notable exceptions. The \xmm\ data set from 2012 (which has the highest signal-to-noise ratio) shows a trend in the residuals around 1 keV, when fitted with a power-law (Fig.~\ref{fig:xmmspec}). This could be caused by emission from hot diffuse gas. We therefore added a second component (MEKAL in \textsc{Xspec}) to the model for this spectrum. This model yields a hydrogen column density of $5.2\pm0.5$ cm$^{-2}$, power-law photon index of $2.3\pm0.3$, and plasma temperature of 1.3$\pm0.2$ keV. It also reduced the fit $\chi^2$ from 218.5 (for 176 degrees of freedom in fitting a single power-law) to 186.5 (for 174 degrees of freedom).  An F-test gives a probability of $10^{-6}$ for obtaining this reduction in $\chi^2$ by chance, suggesting the second component is statistically significant. Testing this model on other spectra that have moderate signal-to-noise levels (e.g., \xmm\ data from 2000-09-21 and \chandra\ data from 2006-06-27) suggests that a MEKAL component is present in those data sets as well (Table~\ref{tab:xmmspec}, Fig.~\ref{fig:xmmspec}). The presence of a low-temperature plasma component, giving detectable X-ray line emission around 1 keV, is typical of CVs \citep[e.g.][]{Baskill05,Wada17}.

\begin{figure}
\includegraphics[scale=0.6]{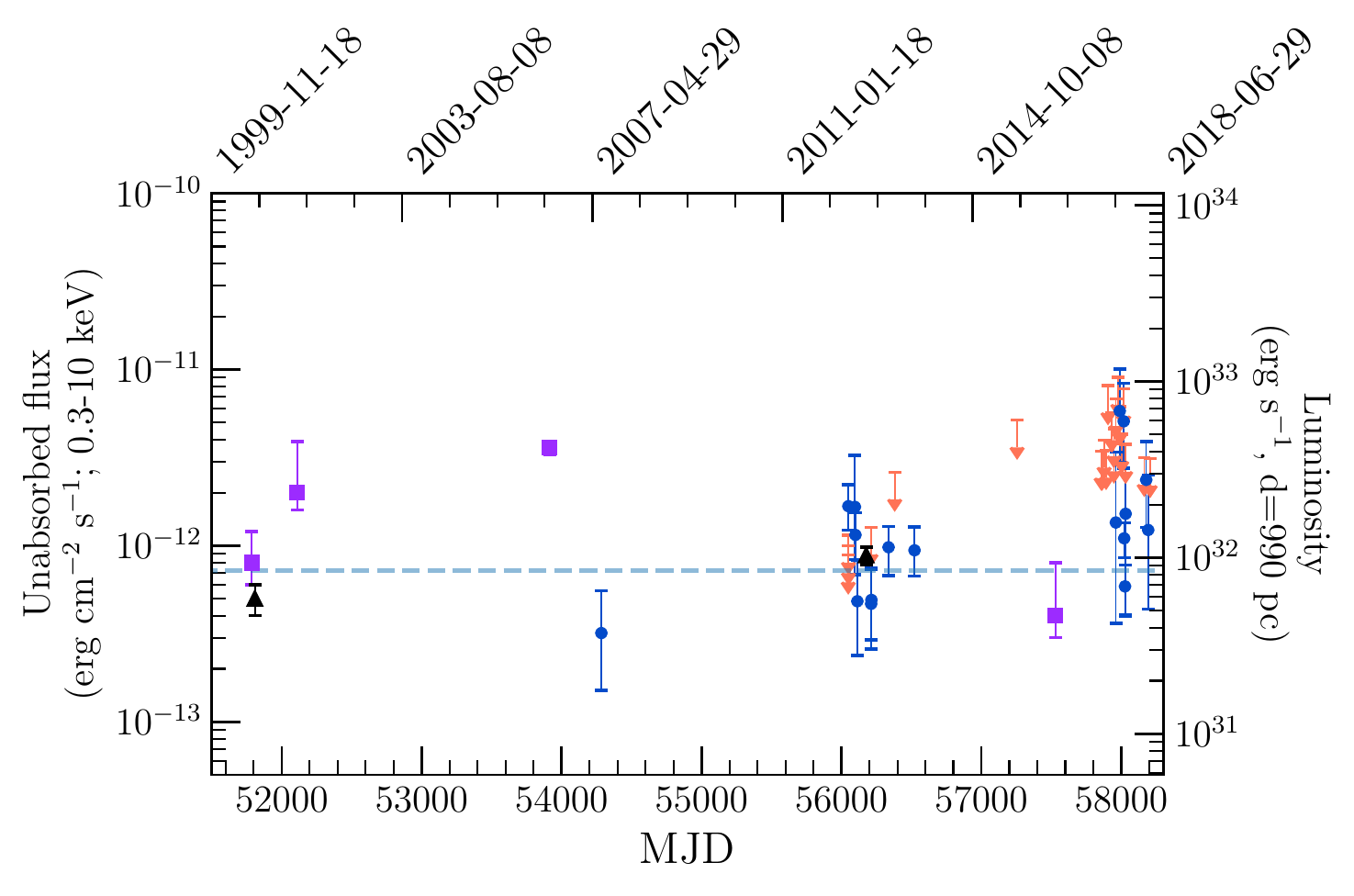}
\caption{X-ray light curve of \srcxmm\ as seen by \xrt\ (blue circles and orange upper limits), \chandra\ (purple squares) and \xmm\ (black triangles). \xrt\ count rates were converted assuming a power-law model with a photon index of 2.5 (as this model describes the source behavior in most cases).}
\label{fig:xmmlc}
\end{figure}

Comparing the source flux as estimated from \xrt\ count rates along with \xmm\ and \chandra\ spectra provides significant evidence that the source is strongly variable (Fig.~\ref{fig:xmmlc}). Assuming a distance of 0.99 kpc (based on the optical counterpart, \citealt{Shaw20}), this indicates that the source is variable between $5\times10^{31}$ and $5\times10^{33}$ \ergs. The average $L_X$ seems to be compatible with the calculated range of $L_X$ for the focused wind CV model of \citet{Shaw20}, $9\times10^{30}-2\times10^{32}$ \ergs.
\begin{table*}
\centering
\caption{Results of spectral analysis for \srcxmm, when fitting an abosrbed power-law (TBABS$\times$PEGPWRLW in \textsc{Xspec}, top panel), and an absorbed power-law + hot plasma (TBABS$\times$[PEGPWRLW+MEKAL] in \textsc{Xspec}, bottom panel). F$_\textrm{X}$ is the unabsorbed flux in the 0.3--10 keV band.}
\label{tab:xmmspec}
\begin{tabular}{lcccccc}
\hline
\hline
Date                        & 2000-08-30                    & 2000-09-21                    & 2001-07-20/21                     & 2006-06-27/28 \& 07-01            & 2012-09-07                        & 2016-05-21                    \\
Telescope                   & \chandra\                     & \xmm\                         & \chandra\                         & \chandra\                         & \xmm\                             & \chandra\                     \\
Obs.ID.                     & 00658                         & 0112970801                    & 02278,02281,02289                       & 07042, 07345, 07346               & 0694640101                        & 18331, 18332, 18333           \\
\hline
N${_\text{H}}$ (cm$^{-2}$)  & $5\pm2\times10^{21}$          & $6\pm1\times10^{21}$          & $8\pm4\times10^{21}$             & $6\pm1\times10^{21}$              & $6.2\pm0.5\times10^{21}$          & $<5\times10^{21}$             \\
Photon index                & $2.5\pm0.4$                   & $2.9\pm0.3$                   & $2.5\pm0.3$                       & $2.7\pm0.2$                       & $2.6\pm0.1$                       & $2.2_{-0.4}^{+0.7}$           \\
Calib. (MOS1/PN)            & --                            & $1.2\pm0.2$                   & --                                & --                                & --                                & --                            \\
Calib. (MOS2/PN)            & --                            & $1.2\pm0.2$                   & --                                & --                                & $0.93\pm0.06$                     & --                            \\
F$_\textrm{X}$ (\ergcms)    & $8_{-2}^{+4}\times10^{-13}$   & $7\pm2\times10^{-13}$         & $23_{-5}^{+8}\times10^{-13}$     & $9\pm2\times10^{-13}$             & $12\pm1\times10^{-13}$            & $4_{-1}^{+4}\times10^{-13}$   \\
$\chi^2_\nu$/D.O.F          & 0.5/11                        & 1.1/45                        & 0.98/74                            & 1.6/18                            & 1.2/176                           & 0.3/2                         \\
N.H.P (\%)                  & 90                            & 27                            & $51$                            & 5.7                               & 1.6                               & 75                            \\
\hline
N${_\text{H}}$ (cm$^{-2}$)  & --                            & $5\pm1\times10^{21}$          & $9_{-7}^{+6}\times10^{21}$       & $14_{-8}^{+4}\times10^{21}$       & $5.2\pm0.5\times10^{21}$          & --                            \\
kT (keV)                    & --                            & $1.1_{-0.2}^{+0.3}$           & $1.2_{-0.7}^{+4.4}$            & $0.3_{-0.1}^{+0.3}$               & $1.3\pm0.2$                       & --                            \\
Norm$_\textrm{MEKAL}$       & --                            & $5_{-2}^{+4}\times10^{-5}$    & $52_{-40}^{+35}\times10^{-5}$  & $185_{-179}^{+1689}\times10^{-5}$ & $11\pm4\times10^{-5}$             & --                            \\
Photon index                & --                            & $2.5\pm{0.4}$                 & $<2.5$                      & $2.8\pm0.3$                       & $2.3\pm0.2$                       & --                            \\
Calib. (MOS1/PN)            & --                            & $1.2\pm0.2$                   & --                                & --                                & --                                & --                            \\
Calib. (MOS2/PN)            & --                            & $1.2\pm0.2$                   & --                                & --                                & $0.93\pm0.06$                     & --                            \\
F$_\textrm{X}$ (\ergcms)    & --                            & $5\pm1\times10^{-13}$         & $20_{-4}^{+19}\times10^{-13}$  & $36\pm3\times10^{-13}$            & $8.8_{-0.6}^{+0.8}\times10^{-13}$ & --                            \\
$\chi^2_\nu$/D.O.F          & --                            & 0.9/43                        & 0.92/72                            & 0.9/16                            & 1.1/174                           & --                            \\
N.H.P (\%)                  & --                            & 56                            & 66                                 & 60                                & 24                                & --                            \\
\hline
F-Test probability(\%)      & --                            & 1.2                           & 3.8                                 & 0.3                               & $8\times10^{-7}$                  & --                            \\
\hline
\end{tabular}
\end{table*}

\section{Discussion}

\subsection{Classes of objects}

We first enumerate the numbers of different classes of objects we have identified so far:
104 sources, minus 13 due to optical loading, gives 91 real X-ray sources. 35 of these sources are substantially (factor $>$5) brighter in at least one SBS epoch, compared to \chandra\ or \xmm\ archival data.  45 are known sources:
\begin{itemize}
\item 12  XRBs
\item 5  CVs
\item 3 symbiotics
\item 4 candidate CVs or XRBs
\item 17 chromospherically active stellar X-ray sources.
\item 1 young star cluster, 2 radio pulsars
\item 1 AGN
\end{itemize}
46 sources of unknown nature:
\begin{itemize}
\item 6 sources with secure optical counterparts which are likely produced by chromospheric activity. These are SBS X 136, 148, 180, 417, 669, 744.
\item 8 sources with secure optical counterparts which are likely accretion-fed (LMXBs, CVs, symbiotics, or AGN) in nature.
\item 11 soft foreground sources not securely associated with stars, but likely produced by chromospheric activity. The lack of a secure optical association may often be due to a lack of a subarcsecond (\chandra) X-ray position. Many of these show dramatic X-ray flaring, up to more than a factor of 100.
\item 13 hard persistent sources, which must be luminous and distant. A large fraction of these, perhaps most, are likely AGN, though other natures--e.g. IPs or X-ray binaries--are possible.
\item 7 hard sources that clearly increased in flux by a factor $>$5.  We suggest the majority of these are very faint X-ray binaries.
\end{itemize}
\subsection{Expected and observed populations of AGN, CVs, ABs}
We can estimate the numbers of several different populations that we expect among our X-ray sources using prior population estimates. We take a rough estimate of our (stacked) X-ray flux sensitivity as $F_X$(2--10 keV)$\sim10^{-12}$ \ergcms. We predict, from \citet{Ueda98}, $\sim0.6$ AGN deg$^{-2}$ at this flux level, thus about 10 AGN in our survey field. 

How can we identify these AGN? AGN (in the direction of the Galactic bulge) will typically be extinguished by $N_H$ of order $10^{22}$ cm$^{-2}$, thus should be relatively hard X-ray sources with the majority of counts above 1.5 keV. All AGN seem to be variable in X-rays, but typical variability in the hard X-rays on month timescales seems to be around 30\%, up to a factor of 3 for the narrow-line Seyfert 1 class, or up to 10 in the more variable soft X-rays \citep{Ulrich97,Gaskell03,Paolillo04}. Observations of $\sim70$ AGN in the \chandra\ Deep Fields found that variation of $>$4 in $L_X$ were very rare \citep{Yang16,Zheng17}. \citet{Strotjohann16} found that 5\% of a sample of AGN observed with both \rosat\ and the \xmm\ Slew Survey varied by a factor of 10 in the soft band. We consider that AGN varying by $>$10 in 2--10 keV will be rare.

Although we only securely identify one AGN in our sample, we suspect that many, perhaps the majority, of the 13 hard persistent sources are likely to be AGN . This would be consistent with the predictions above.

We expect some cataclysmic variables within our survey field. \citet{Jonker11} predicts $\sim$5 IPs at our flux levels within 16 square degrees, which should have optical counterparts of $r\sim17-19$ \citep{Britt14} if seen under relatively low extinction, $N_H<3\times10^{21}$ cm$^{-2}$. The brightest IPs (at $L_X=10^{34}$ \ergs) may be visible at Galactic Centre distances. Indeed, we find 5 known CVs in our field, plus another 4 candidate CVs or XRBs.

Chromospherically active binaries will generally be lower-$L_X$, and visible at shorter distances. \citet{Jonker11} suggest $10^{-4}$ RS CVn binaries pc$^{-3}$, which with a typical $L_X=10^{31}$ \ergs, suggests a maximum distance of 300 pc, with about 4 in our survey (at $r\sim10$, \citealt{Britt14}). The population synthesis of \citet{Warwick14} predicts the number of X-ray detected ABs in Galactic plane fields at $F_X$(2--10 keV)$\sim10^{-12}$ \ergcms\ to be 1/2 the number of CVs, and 1/6 the number of XRBs, suggesting the number of detected ABs should be $\lesssim$10 (based on the numbers of other categories). However, we find 17 secure ABs, 12 foreground sources that are probably mostly or all ABs, as well as 13 sources with optical counterparts of uncertain nature, suggesting a total of 30-40 ABs in our sample.

The population syntheses mentioned do not consider flares, which we appear to be seeing in abundance, as 2/3 of the known chromospherically active stars in our sample with prior \chandra\ or \xmm\ observations were seen to be flaring by a factor of 5 or more. Soft foreground sources not identified with stars tended to have even stronger flares; 7 of 10 with prior \chandra/\xmm\ imaging flared by a factor of 10 or more, with three flaring by over a factor of 100. We interpret this as a selection effect; stars that flare from initially lower X-ray fluxes up to our flux limit may not be detected by \chandra\ in short observations in quiescence, and thus are harder to confidently identify with their optical counterparts (as their error radii are larger).

\subsection{Predicted population of VXFTs in the bulge}
The \swift\ Galactic Bulge Survey was inspired in large part by the identifications of a number of VFXTs in the long-running \xrt\ survey of the Galactic Centre \citep{Degenaar09,Degenaar10,Degenaar13,Degenaar15}. 
Here we assume that X-ray binaries are distributed uniformly in stellar mass, calculate the expected number of both bright and fainter (VFXT) transients that we might detect in our survey based on the Galactic Centre studies above, and compare it to our actual results, to see if the Galactic Centre is unusual. The Galactic Centre \swift\ survey largely used daily 1-ks observations, allowing high-quality light curves of each detected X-ray outburst. We used PIMMS\footnote{\url{http://asc.harvard.edu/toolkit/pimms.jsp}} to calculate the 2--10 keV X-ray luminosities of each observation of an outburst within the first 4 years of the \swift\ program \citep{Degenaar10}, along with the peak and average L$_X$ values. We separated the outbursts into three classes: Bright (peak $>10^{36}$ \ergs), Very Faint A (peak $10^{35} - 10^{36}$ \ergs), and Very Faint B (peak $<10^{35}$ \ergs). Note that some LMXBs exhibit two, or all three, types of outburst at different times \citep{Degenaar09,Degenaar10}. We used the Galactic mass model of \citet{Launhardt02} to estimate the mass covered by the Galactic Centre \swift\  pointings, and thus calculated the rate of each type of outburst per stellar mass. 

We then modeled each of the 163 pointings of SBS (in the 2017 season) to predict the number of outbursts expected to be detectable from our survey. We created fake light curves corresponding to actual light curves of known Galactic Centre transients at random times, distributed uniformly by Galactic stellar mass (using the Launhardt model), with count rates calculated using the model L$_X$ at that point in the outburst, and the N$_{\mathrm{H}}$ estimated by \citet{Schlegel98} in that direction. We performed $10^4$ simulations of the SBS survey.

We used this model to test the expected numbers of detections of bright, `very faint A', and `very faint B' outbursts expected with our observational setup, assuming straightforward scaling of X-ray binaries by stellar mass. The median results were a prediction of 24 bright outbursts, 4.7 `very faint A' outbursts, and 0.9 `very faint B' outbursts. Comparing this to our observations, we see that of the 12 SBS sources associated with known LMXBs, we find 8 to be bright ($L_X>10^{36}$ \ergs), 6 of which are persistent through our observations; 3 `very faint A' outbursts ($10^{35}<L_X<10^{36}$ \ergs); and 1 `very faint B' outburst ($L_X<10^{35}$ \ergs, from GRO J1744$-$28). However, we also see seven hard transient sources, which may be  `very faint B' LMXB outbursts, though there are other possibilities. Overall, we find that our modeling reasonably represents our detection rates of `very faint A' and `very faint B' outbursts, but predicts too many bright outbursts, compared to what we actually see. This could indicate an excess of bright X-ray binaries at the Galactic Centre, compared to the Bulge regions we are studying here.

We also modeled different observing strategies to attempt to increase the number of `very faint A' and `very faint B' outbursts that can be detected with a similar observation time (relevant for planning future surveys). We found that a strategy of spacing the pointings farther apart to minimize overlaps, while increasing the time per pointing from 60 to 120 seconds (thus from 45 to 105 seconds average exposure, accounting for slew and settle times), gave significantly higher probabilities of detecting `very faint B' transients, and of detecting `very faint A' transients during the tails of their outbursts (as they tend to spend significantly longer with $L_X < 10^{35}$ \ergs\ than above it).  Using a biweekly observation scheme with 142 pointings, rather than 163, over 19 epochs predicts a total of 14 bright outburst detections (similar to that predicted for the 2017 season), but with 6-7 `very faint A', and 1.4 `very faint B' detections, predicted.  The very faint numbers are 40-50\% more than obtained with the 2017 60-second exposure plans as listed above (though these small numbers would be subject to Poisson fluctuations). We tested this new deeper-exposure plan in the last few cycles of the 2017-18 \swift\ Bulge Survey, in Feb-March 2018, and encountered no problems. We implemented this new system in our 2019 observations of the SBS.

\subsection{Symbiotic systems}
Finally, we note that we have identified a substantial number of possible symbiotic systems. We use the terminology ``symbiotic star'' to indicate a white dwarf accretor,  ``symbiotic X-ray binary'' for a neutron star or black hole accretor, and ``symbiotic system'' where the accretor nature is uncertain. In our survey, these include the likely symbiotic X-ray binaries SBS X4 and SBS X9, the hybrid CV/symbiotic star SBS X374, and the strong candidate symbiotic systems SBS X707, X97, X109, X557, and X771.  Symbiotic stars may also be present among the remaining 10 sources with optical counterparts of unknown nature, the 7 hard transients, and possibly the 13 hard persistent systems.

Existing models of the symbiotic star population in the Galaxy (e.g. \citealt{Lu06}) do not estimate the X-ray luminosity of symbiotic stars. However, if we take the \citet{Lu06} estimate of the total number of symbiotic stars in the Galaxy ($10^3$--$10^4$), and use the local stellar density (0.1 stars pc$^{-3}$) and 200 billion stars in the Galaxy, we can estimate a space density of $5\times10^{-8}$--$5\times10^{-9}$ symbiotic stars pc$^{-3}$. If we assume a peak X-ray luminosity of $10^{33}$ \ergs (and thus a maximum distance of $\sim$3 kpc, as our faintest SBS sources are around $10^{-12}$ \ergcms), then our 16 square degrees gives us a volume of $4.4\times10^7$ pc$^3$, and $<$0.2$f$ symbiotic stars predicted (where $f$ is the fraction that reach this $L_X$).  At $10^{34}$ \ergs, we are sensitive to objects in the denser Bulge, at least in the less extinguished regions; this gives us access to $\sim$6\% of the stars in the Galaxy \citep{Jonker11}. Symbiotic stars are not thought to reach $10^{34}$ \ergs\ in hard X-rays, except during rare nova explosions \citep{Mukai08}, so it seems unlikely to us that symbiotic stars in the Bulge contribute signficantly. A larger population of X-ray detectable symbiotic stars at lower accretion rate has recently been suggested \citep{vandenBerg06,Hynes14,Mukai16}, based on the idea that accretion rates below $10^{-9}$ \Msun/yr can give high X-ray efficiency without reaching the bolometric accretion luminosity of 10 \Lsun\ necessary for optical identification of a symbiotic star. A substantial number of symbiotic stars at distances 1-3 kpc in our survey would give strong support for this hypothesis.

Alternatively, symbiotic X-ray binaries can  provide $L_X\sim10^{34}$ \ergs\ or more. \citet{Lu12} predict a total Galactic population of 100--1000 symbiotic X-ray binaries. \citet{Lu12} predict most to have $L_X<10^{33}$ \ergs, but do not study whether this accretion is steady, or may occur in (wind clump-induced) flares. UV observations (e.g. Rivera~Sandoval et al. in prep.) will be especially helpful in investigating potential nearby symbiotic systems, as  accreting white dwarfs should produce a substantial UV excess. 

\section{Conclusion}
The \swift\ Bulge Survey covered 16 square degrees of the Galactic Bulge in 19 epochs, using 60-second \xrt\ exposures. The data were taken at 2-week intervals during April-Oct 2017, and Feb-March 2018. This allows us to probe faint transient X-ray behaviour, at fluxes below those detectable by all-sky monitors, across a significant fraction of our galaxy.
We identify 104 clear \xrt\ detections, 13 of which we attribute to optical loading of the detectors, and thus 91 significant X-ray sources. 25 of these sources varied by a factor of 10 or more (generally in comparison to archival Chandra or XMM data), and another 14 by at least a factor of 5. Since many of our Swift detections had large $L_X$ error bars, and not all sources were covered by archival data, the true fraction of our sample that are transients is probably significantly higher.

We can identify 45 of these sources as known categories of X-ray sources: X-ray binaries, cataclysmic variables, symbiotic systems, chromospherically active stars, radio pulsars, an AGN, and a young star cluster. We include detailed X-ray spectroscopy of several objects of particular interest, including the bursting X-ray transient \srcigr\ (which we and \citealt{Shaw20} argue is a symbiotic X-ray binary with a low $B$ field), and a subgiant star in a 8.7 day orbital period with a white dwarf (which we and \citealt{Shaw20} argue is a focused-wind system, in between a CV and a symbiotic star).

Another 46 sources do not have certain classifications yet. We use their hardness ratios, partial information about their optical/infrared counterparts, and measurements of how much their flux has increased compared to quiescence (\chandra\ or \xmm\ data), to suggest that these include a mixture of AGN, chromospherically active stars, symbiotic stars, and LMXBs undergoing very faint outbursts. Our survey found 4 detections of known LMXBs when $L_X<10^{36}$ \ergs, proof that we are sensitive to VFXT behaviour. We have also found at least 7 unknown hard X-ray sources that showed outbursts by at least a factor of 5, without obvious optical/infrared counterparts, suggestive of distant, luminous X-ray transients in our Galaxy.  Our 2019 run of the SBS, with longer (120 s) exposures, detected more very faint transients (which will be described elsewhere).

Several hard X-ray transients with bright optical/infrared counterparts, often at several kpc distances, may be symbiotic systems. Detailed studies of these systems can test this hypothesis. 

In addition to the \xrt\ survey, the SBS survey also provided a survey in the direction of the Galactic Bulge with the UVOT telescope in several UV filters, enabling the detection of UV transients (\sbsuvot).

\section*{Data Availability}
All the data presented/used in this work are publicly available. All the \swift\ data are publicly accessible through the HEASARC portal (\url{https://heasarc.gsfc.nasa.gov/db-perl/W3Browse/w3browse.pl}) and the UK \swift\ Science Data Centre (\url{https://www.swift.ac.uk/archive/index.php}). All the software packages used in this work are open-source and publicly available (see below).

\section*{Acknowledgements}
We thank the anonymous referee for their helpful suggestions. AB thanks John Good for helpful discussions and James C.~A. Miller-Jones for support during this project. TJM and LERS thank NASA for support under grant 80NSSC17K0334. COH, GRS and AJT acknowledge support from Natural Sciences and Engineering Research Council of Canada (NSERC) Discovery Grants (RGPIN-2016-04602 and RGPIN-06569-2016 respectively), and a Discovery Accelerator Supplement for COH. ND is supported by a Vidi grant from the Netherlands Organization for Scientific Research (NWO). PAE acknowleges UKSA support. AJT acknowledges support for this work through an NSERC Post-Graduate Doctoral Scholarship (PGSD2-490318-2016). JS acknowledges support from a Packard Fellowship. We acknowledge extensive use of NASA's Astrophysics Data System Bibliographic Services, Arxiv, SIMBAD \citep{Wenger00} and Vizier \citep{Ochsenbein00}.

We have made use of available archival data from \chandra, \xmm\ and \swift.

This research has used the following software packages: Astropy \citep{Astropy13,Astropy18}, BXA \citep{Buchner14}, HEASOFT \citep{HEASARC14}, IPython \citep{Perez07}, Jupyter \citep{Kluyver16}, Matplotlib \citep{Hunter07}, MultiNest \citep{Feroz19}, Numpy \citep{oliphant06,vanderwal011}, SAOImage DS9 \citep{Joye03}, XSPEC \citep{Arnaud96}.




\bibliographystyle{mnras}
\interlinepenalty=10000
\bibliography{full_bibliography} 



\bsp	
\label{lastpage}
\end{document}